\documentclass[aps,prl,superscriptaddress,twocolumn,floatfix]{revtex4-1}
\usepackage{CJK}
\usepackage{graphicx,amsmath,amssymb,color,multirow,xcolor, MnSymbol,hyperref}
\usepackage{soul}
\usepackage{todonotes}

\begin{document}

\begin{CJK}{UTF8}{mj}
\title{Incorporating Heterogeneous Interactions for Ecological Biodiversity}
\author{Jong Il Park (박종일)}
\affiliation{Department of Physics, Inha University, Incheon 22212, Korea}
\author{Deok-Sun Lee (이덕선)}\email[Corresponding author: ]{deoksunlee@kias.re.kr}
\affiliation{School of Computational Sciences, Korea Institute for Advanced Study, Seoul 02455, Korea} 
\author{Sang Hoon Lee (이상훈)}\email[Corresponding author: ]{lshlj82@gnu.ac.kr}
\affiliation{\mbox{Department of Physics and Research Institute of Natural Science, Gyeongsang National University, Jinju 52828, Korea}} 
\affiliation{\mbox{Future Convergence Technology Research Institute, Gyeongsang National University, Jinju 52849, Korea}}
\author{Hye Jin Park (박혜진)}\email[Corresponding author: ]{hyejin.park@inha.ac.kr}
\affiliation{Department of Physics, Inha University, Incheon 22212, Korea}

\date{\today}
\begin{abstract}
Understanding the behaviors of ecological systems is challenging given their multi-faceted complexity. 
To proceed, theoretical models such as Lotka-Volterra dynamics with random interactions have been investigated by the dynamical mean-field theory to provide insights into underlying principles such as how biodiversity and stability depend on the randomness in interaction strength. 
Yet the fully-connected structure assumed in these previous studies is not realistic as revealed by a vast amount of empirical data. 
We derive a generic formula for the abundance distribution under an arbitrary distribution of degree, the number of interacting neighbors, which leads to degree-dependent abundance patterns of species.
Notably, in contrast to the well-mixed system, the number of surviving species can be reduced as the community becomes cooperative in heterogeneous interaction structures.
Our study, therefore, demonstrates that properly taking into account heterogeneity in the interspecific interaction structure is indispensable to understanding the diversity in large ecosystems, and our general theoretical framework can apply to a much wider range of interacting many-body systems.
\end{abstract}

\maketitle

%Intro--------
{\it Introduction.---}
The introduction of the field-theoretic approach~\cite{martin1973statistical,janssen1976lagrangean,de1978field,tauber2014critical} for stochastic processes has led to its application across various dynamical systems, including neural networks~\cite{crisanti2018path,sompolinsky1988chaos,helias2020elementary}, statistical learning~\cite{mannelli2020marvels,poole2016exponential,segadlo2022unified}, and game theory~\cite{galla2009minority,baron2021consensus,opper1992phase}. 
Dynamical mean-field theory (DMFT), initially devised for investigating spin-glass dynamics~\cite{sompolinsky1981dynamic,crisanti1993spherical}, has found success in ecological modeling, particularly in solving generalized random Lotka-Volterra (GRLV) equations of species abundance~\cite{tikhonov2017collective,yang2023enhancing,galla2018dynamically,bunin2017ecological}.
This breakthrough sheds light on how complexity impedes the stability and diversity of large ecological communities~\cite{galla2018dynamically,bunin2017ecological,roy2019numerical,biroli2018marginally}, aligning with earlier findings from the random matrix theory~\cite{may1972will,allesina2012stability, baron2023breakdown}.

While the current DMFT method is predominantly constrained to fully-interacting or well-mixed communities, real-world data reveal complicated and heterogeneous structures in ecological networks~\cite{jordano2003invariant,krause2003compartments,dunne2002network,guimaraes2020structure,montoya2002small}, making them archetypical examples of complex networks in network science.
From the perspective of network science, understanding the impact of structural heterogeneity on dynamics stands as a central issue, given its universality across disciplines~\cite{NewmanBook,albert2002statistical,barrat2008dynamical}. 
In response, theoretical tools such as the heterogeneous mean-field theory (HMFT) have been developed~\cite{dorogovtsev2008critical,pastor2001epidemic}. 
The HMFT, based on the expectation that agents with the same number of interacting partners exhibit identical dynamic behaviors, has successfully explained the critical phenomena in spin systems of a general network structure, offering insights distinct from the conventional mean-field results~\cite{kim2001xy,dorogovtsev2002ising,igloi2002first,kim2005spin,lee2009critical}.
Moreover, the HMFT yielded fruitful insights into epidemics~\cite{castellano2010thresholds,costa2022heterogeneous} and synchronization~\cite{lee2005synchronization,rodrigues2016kuramoto} across various networks. 

In Ref.~\cite{barbier2018generic}, the DMFT has been tested for its performance in explaining ecological systems on complex networks by extracting the interaction mean and variance at the system level.
Although this approach effectively describes ecological systems with relatively homogeneous interaction structures, it fails to provide accurate approximations for heterogeneous cases. 
As an endeavor to incorporate structural heterogeneity into the theoretical framework for the dynamics of ecological systems, solvable models beyond the well-mixed structures have been investigated~\cite{lee2022stability,marcus2022local,poley2023generalized}. 
However, developing a general framework that can account for both structural heterogeneity and randomness in interaction strength remains a fundamental challenge.
Such a framework is essential for comprehensively understanding the diversity and stability of ecological systems, but it has yet to be done.

In this Letter, we propose a generic theoretical framework, {\it heterogeneous dynamical mean-field theory} (HDMFT), combining the two theoretical methods, the DMFT from the field-theoretic approach and the HMFT from network science, and apply it to the GRLV dynamics on general ecological networks and obtain the species abundance distribution and survival probability at the system and individual level.
The most remarkable finding is that the number of surviving species diminishes as the community becomes cooperative in heterogeneous interaction structures, which is counterintuitive and never seen in well-mixed systems. 
The origin lies in the differentiated abundance and survival of individual species with their numbers of interacting partners, and our detailed analysis reveals the interplay between the heterogeneous structure and random strength of interaction.  
This combined framework deepens the understanding of heterogeneous ecological systems and can be utilized in various interacting systems beyond the scope of natural ecosystems.
\newline
%%%%%%%%%%

\begin{figure}%[h!]
    \centering
    \includegraphics[width=\linewidth]{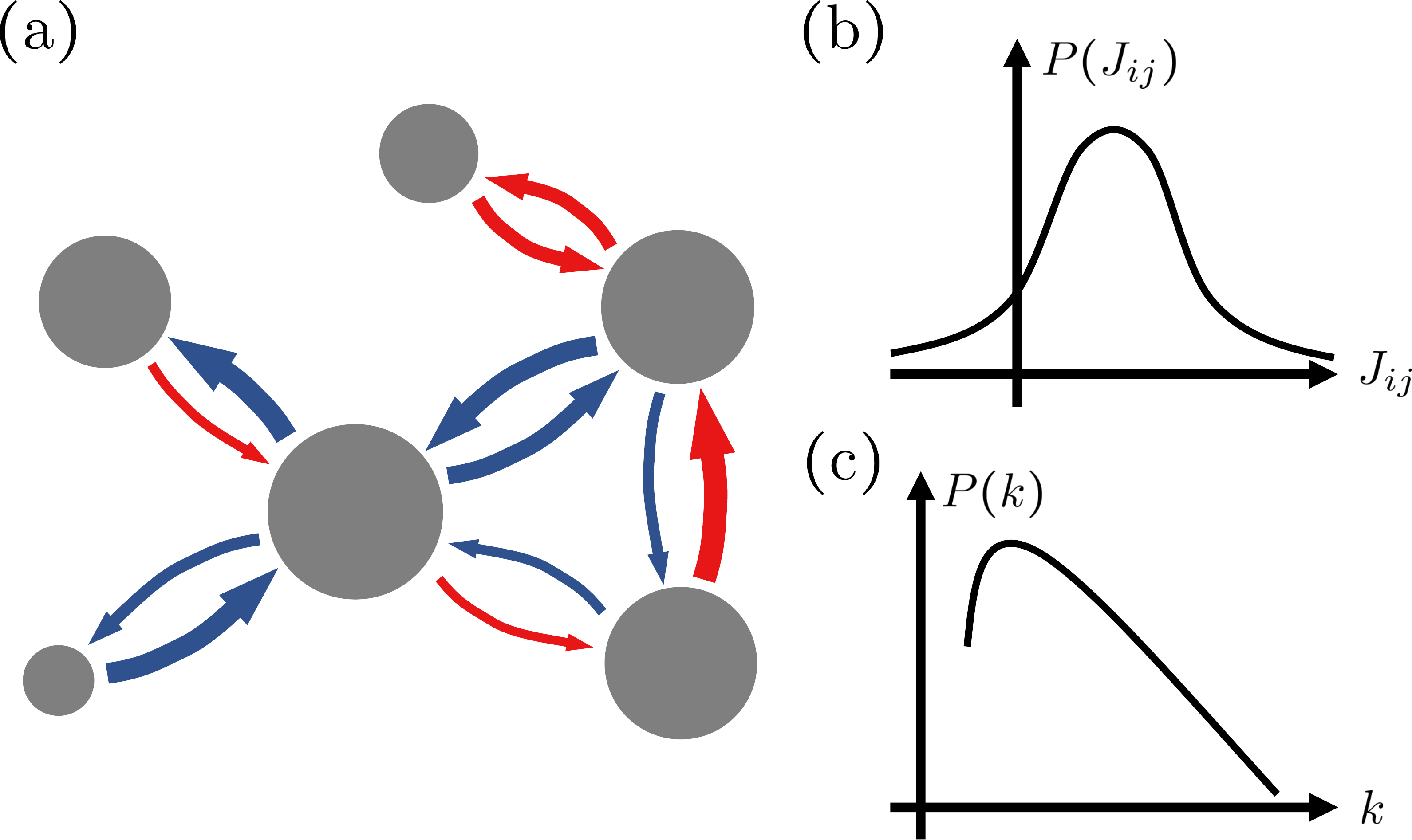}
    \caption{Heterogeneous structure and random strength of interaction. 
    (a) An ecological network of 6 nodes (species) with a directed link from node $j$ to $i$ representing the positive (red) or negative (blue) influence $J_{ij}$ of species $j$ on the growth of $i$ with strength represented by the line width.
    Assuming mutual influences, we assign either none or two opposite directional links to each pair of nodes. 
    (b) The strength $J_{ij}$ follows a Gaussian distribution. 
    (c) Species may have different degrees $k$, numbers of interacting neighbors, following a broad distribution $P(k)$.}
    \label{fig:fig1}
\end{figure}

%Model---------
{\it Model.---}
We first construct a GRLV system with $S$ species as follows:
\begin{align}
    \dot{x}_i(t) = x_i(t) \left[ \lambda - x_i(t) - \sum_{j \backslash i} J_{ij}A_{ij}x_j(t)\right],
    \label{eq:LV}
\end{align}
where $x_i$ is the abundance of $i-$th species with a self-growth rate $\lambda$, and $\sum_{j\backslash i}$ denotes the summation over all species except $i$.
The last term represents the interspecific interaction with two factors: (i) the strength of interaction $J_{ij}$ depicting how much a species $j$ affects another species $i$ positively or negatively and (ii) the structure of interaction (the adjacency matrix in the context of networks) $A_{ij}=1$ or $0$ indicating the existence or absence of interaction (see Fig.~\ref{fig:fig1} for a graphical illustration).
A positive (negative) value of $J_{ij}$ indicates the suppression (promotion) of the growth of $i$ by $j$.
We set the adjacency matrix to be symmetric, i.e., $A_{ij}=A_{ji}$ while we consider $J_{ij}$ and $J_{ji}$ as independent or correlated to some extent controlled by a parameter.

These interspecific interactions, $J_{ij}$ and $A_{ij}$, do not need to be uniform in nature, and thus we assume the randomness in both $J_{ij}$ and $A_{ij}$. 
As in previous studies~\cite{bunin2017ecological,galla2018dynamically,roy2019numerical}, for analytic treatment, we set the interaction strength $J_{ij}$ as a Gaussian random variable with the first few moments given by 
\begin{align}
    {\langle J_{ij} \rangle}_\mathbf{J} = \frac{\mu}{K},\quad {\llangle J_{ij}^2 \rrangle}_\mathbf{J} = \frac{\sigma^2}{K}, \textrm{ and } {\llangle J_{ij} J_{ji} \rrangle}_\mathbf{J} = \frac{r\sigma^2}{K},
\end{align}
where $K = S^{-1}\sum_{i,j} A_{ij}$.
Notations ${\langle \cdots \rangle}_\mathbf{J}$ and ${\llangle \cdots \rrangle}_\mathbf{J}$ represent moment and cumulant, respectively, averaged over an ensemble of interaction strength realizations $\{\mathbf{J}_1,\mathbf{J}_2,\cdots\}$.
The sign of $\mu$ represents whether the considered community is overall competitive $(\mu >0)$ or cooperative $(\mu<0)$. The $\sigma$ characterizes the randomness of interaction strength.
The reciprocity $-1\leq r \leq 1$ is related to the type of pairwise interactions, e.g., $r=-1$ with predator-prey interactions $(J_{ij} = -J_{ji})$.

To represent a heterogeneous interaction structure, we consider an ensemble of adjacency matrices $\{\mathbf{A}_1, \mathbf{A}_2,\cdots\}$ with each element $A_{ij}$ taking $0$ and $1$ with probability $1-p_{ij}$ and $p_{ij}$ respectively, and thus satisfying
\begin{align}
    {\langle A_{ij} \rangle}_\mathbf{A} = {\langle A_{ij}^2 \rangle}_\mathbf{A} = \cdots = p_{ij} \,.
\end{align}
The number of interacting partners of species $i$ is given by $k_i = \sum_j A_{ij}$ called {\it degree} and its ensemble average $\langle k_i\rangle = \sum_{j \backslash i} p_{ij}$ can be heterogeneous if the connection probabilities $p_{ij}$ are not identical. 
We use $p_{ij}$ proposed in the static model~\cite{goh2001universal} to generate networks with the power-law degree distributions $P(k)\sim k^{-\gamma}$, where $\gamma$ is the degree exponent, or the Poisson distribution $P(k) = \mathrm{Pois}(k;K) = K^ke^{-K}/{k!}$~\cite{Erdos1959,Gilbert1959}.
We here approximate the connection probability by $p_{ij} \approx k_ik_j/(SK)$, which is often called the annealed approximation and valid in uncorrelated networks without a significant degree-degree correlation~\cite{Newman2002,dorogovtsev2008critical,park2004statistical}.
Under this approximation, the statistical property of $A_{ij}$ is solely determined by the degree sequence $\{k_i\}$, leading to the dynamical equations for the HMFT.
\newline

%MF Results---------
{\it HDMFT.---}
Using the DMFT technique with the HMFT for Eq.~\eqref{eq:LV} and setting $r=0$ for simplicity, one can obtain the abundance dynamics for a species with degree $k$ as follows:
\begin{equation}
\begin{aligned}
    \dot{x}(t;k) = x(t;k)\left[\lambda - x(t;k) - \mu \frac{k}{K}m(t) - \sigma\sqrt{\frac{k}{K}}\eta(t) \right].
    \label{eq:MF}
\end{aligned}
\end{equation} 
The outcome can be roughly understood by approximating $\sum_{j \backslash i} J_{ij} A_{ij} x_j(t)$ in Eq.~\eqref{eq:LV} as the sum of $k_i$ independent and identically distributed random variables, resulting in $\mu \frac{k}{K}m(t)$ along with the Gaussian noise $\eta(t)$, where $\langle \eta(t) \rangle=0$ and $\langle \eta(t) \eta(t') \rangle= q(t, t')$. Here, $m(t)$ and $q(t,t')$ are 
\begin{equation}
\begin{aligned}
    m(t) &= \left\langle \frac{k}{K} x(t;k)\right\rangle,\\
    q(t,t') &= \left\langle \frac{k}{K} x(t;k)x(t';k)\right\rangle.
    \label{eq:OP}
\end{aligned}
\end{equation}
Equations~\eqref{eq:MF} and \eqref{eq:OP} are the main results from the HDMFT and their rigorous derivation with moment-generating functional for general $r$ is given in Supplemental Material (SM) Sec. I.A.~\cite{supp}.
Note that $\langle \cdots \rangle$ is the average over both species and ensembles of $(\mathbf{J},\mathbf{A})$. 

The equilibrium abundance $x(k) \equiv \lim_{t\to\infty} x(t;k)$ follows a truncated Gaussian distribution $\rho(x|k)$, 
\begin{align}
    x(k) = \max\left(0,\lambda - \mu mk/K - \sigma\sqrt{q k/K} z\right)\sim\rho(x|k) \,,
    \label{eq:CondProb}
\end{align}
where $z$ is a random variable sampled from the standard Gaussian distribution $\mathcal{N}(0,1)$.
Compared with the well-mixed system~\cite{bunin2017ecological,galla2018dynamically}, differences are found in the terms proportional to $k$ and $\sqrt{k}$ in Eq.~\eqref{eq:CondProb}, which bring the degree dependence and fundamental changes to the abundance distribution and survival probability.

\begin{figure}
    \centering
    \includegraphics[width=\linewidth]{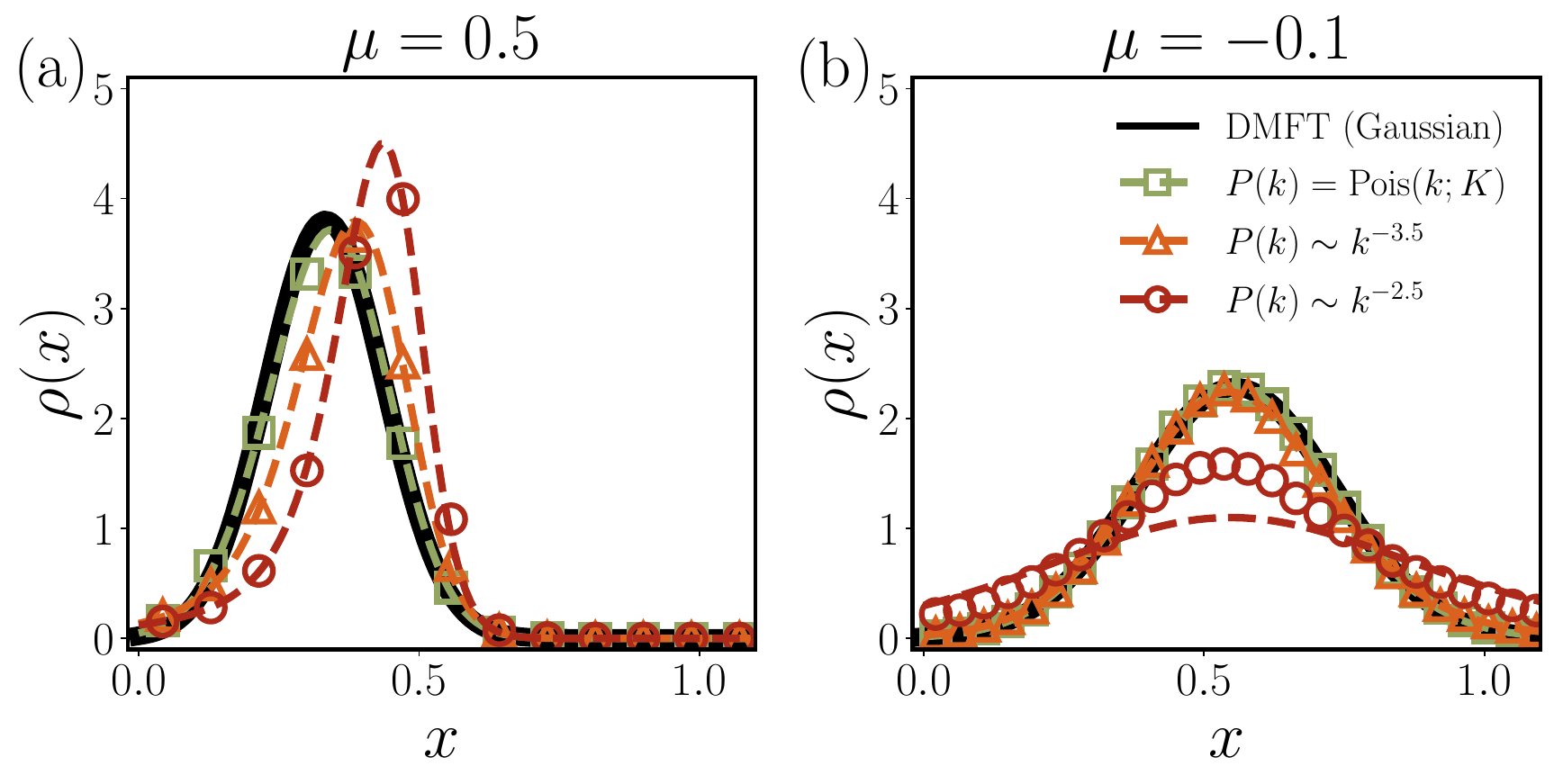} 
    \caption{Species abundance distribution $\rho(x)$ for different degree distributions $P(k)$---the Poisson and power-law distributions with exponents $3.5$ and $2.5$---in (a) a competitive community $(\mu = 0.5)$ and (b) a cooperative community $(\mu = -0.1)$. 
    The thick black lines are obtained from the DMFT approximation without considering structural heterogeneity.
    The dashed curves are obtained from the HDMFT prediction, Eq.~\eqref{eq:CondProb}, and the symbols represent the simulation results averaged over $100$ configurations of $\{\mathbf{J}_\alpha,\mathbf{A}_\alpha\}$. We use the following parameter values: $S = 4000$, $K = 30$, $\lambda = 0.5$, $\sigma = 0.3$, and $r=0$. Unless otherwise noted, we used the same parameters throughout the Letter.
    }
    \label{fig:fig2}
\end{figure}

If the degrees of individual species follow a degree distribution $P(k)$, the stationary abundance distribution of the whole community is given by $\rho(x) = \sum_k P(k) \rho(x|k)$. 
When all the species have the same degree $K$, i.e., $P(k)=\delta_{k,K}$, the truncated Gaussian distribution is recovered, $\rho(x) = \rho(x|K)$ as in fully-interacting systems~\cite{bunin2017ecological,galla2018dynamically}. 
However, when species exhibit varying degrees, the coalition no longer guarantees a truncated Gaussian distribution for $\rho(x)$.
In Fig.~\ref{fig:fig2}, we present $\rho(x)$ obtained with the Poisson distribution $\mathrm{Pois}(k;K)$ and power-law distributions $P(k)\sim k^{-\gamma}$ with $\gamma = 3.5$ and $2.5$, ordered by distribution width (the second cumulant). 
The DMFT approach extracts only the mean and variance of the overall interactions, including zero components, and approximates $J_{ij}A_{ij}$ as a Gaussian random variable with those mean and variance~\cite{barbier2018generic}.
While this method performs well in predicting abundance distributions when degree heterogeneity is relatively small, it fails in cases of large degree heterogeneity.
In contrast, our HMDFT method agrees well with simulation results, even with significant degree heterogeneity. 
Furthermore, our approach accurately describes $\rho(x|k)$, distinguishing between species with varying numbers of interacting species, a capability inaccessible to the DMFT method alone (See Sec. II.C. in SM~\cite{supp}). 

\begin{figure}
    \centering
    \includegraphics[width=\linewidth]{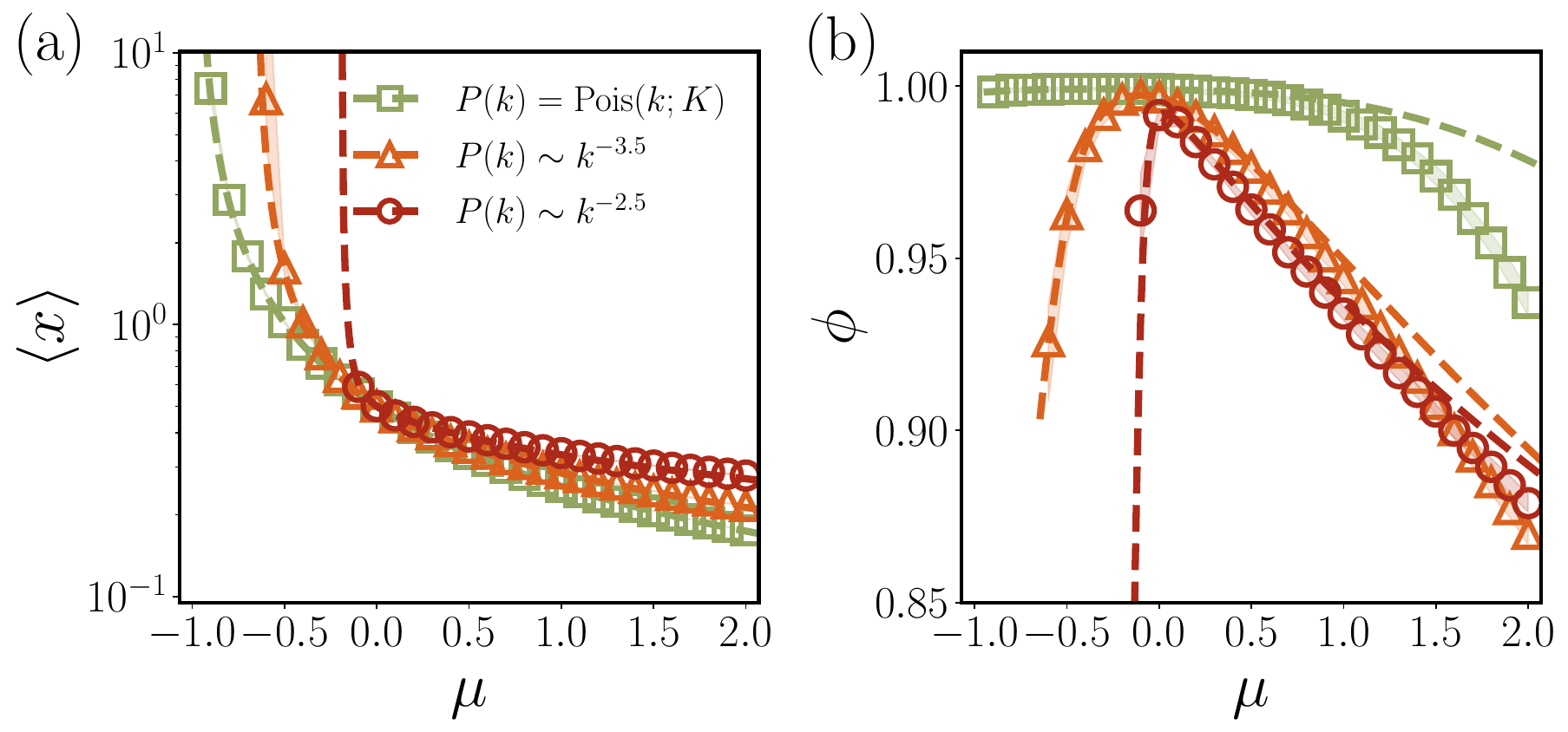}
    \caption{Plots of (a) the average abundance $\langle x \rangle$ and (b) the survival probability $\phi = \langle \Theta(x) \rangle$ against the mean interaction strength $\mu$.
    The dashed curves are theoretical predictions and the symbols represent the simulation results.
    As $\mu$ increases, from negative to positive, the average abundance decreases monotonically for all degree distributions. The survival probability $\phi$ shows non-monotonic behaviors depending on the degree distribution. 
    Particularly, for cooperative communities $(\mu<0)$, $\phi$ decreases even though the system becomes more cooperative on average, notably so for the power-law degree distributions.
    }
    \label{fig:fig3}
\end{figure}

{\it Average abundance and survival probability.---}The system-level influence of structural heterogeneity is best manifested in the average abundance $\langle x \rangle$ and the survival probability $\phi = \langle \Theta(x) \rangle$ with the Heaviside step function $\Theta(x)$.
As shown in Fig.~\ref{fig:fig3}, for a competitive community $(\mu>0)$, both $\langle x \rangle$ and $\phi$ decrease monotonically with $\mu$, i.e., competition curbs the community from growing.
On the other hand, in the cooperative community $(\mu<0)$, as $|\mu|$ gets larger, $\langle x \rangle$ increases but $\phi$ decreases. 
It is counterintuitive, as this result indicates that the more species benefit from each other, the more they eventually vanish, whereas the community itself proliferates.
This dramatic drop in diversity reflected by $\phi$ is absent in well-mixed systems such as fully-connected $(K=S-1)$ case or relatively homogeneous interactions $[P(k) = \mathrm{Pois}(k;K)]$, indicating the important role of interaction structures. 
Without the explicit consideration of degree heterogeneity, the DMFT alone can never predict this phenomenon.

To excavate the reason behind this seemingly counterintuitive behavior appearing in the cooperative community, let us consider the case with no randomness in strength, $\sigma = 0$.
Then all species survive with nonzero abundance $x(k) = \lambda + |\mu|mk/K$, trivially leading to $\phi = 1$.
Thus, cooperative communities are feasible with homogeneous interaction strength by prohibiting survival probability from $\phi$ falling off.
Next, let us consider the case without structural heterogeneity, i.e., $k_i=K$ for all $i$. 
Then one can find that both $m$ and $q$ increase with increasing $|\mu|$ but their ratio $q/m^2$ remains constant and so does the survival probability $\phi$, independent of $\mu$ (See Sec. I.C. in SM~\cite{supp}).
For small $\sigma$, we found that $\gamma<4$ is necessary to observe such a diversity drop (See Sec. I.G. in SM~\cite{supp}).
Therefore, without either heterogeneity in strength or structure is there no diversity drop in $\phi$ for $\mu<0$.

\begin{figure}
    \centering
    \includegraphics[width=\linewidth]{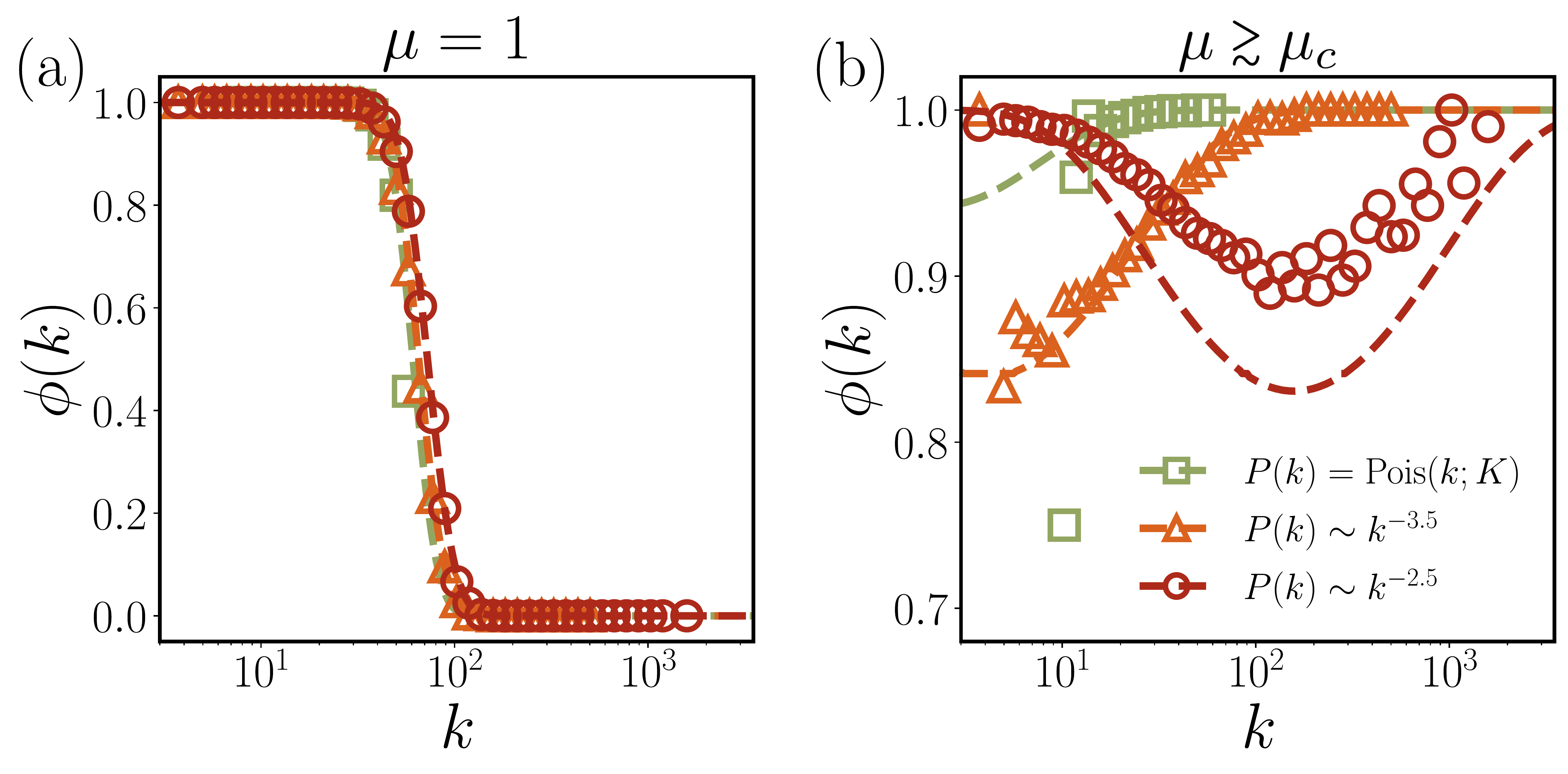}
    \caption{Survival probability $\phi(k)$ of individual species with degree $k$. 
    (a) In competitive communities $(\mu = 1)$, the survival probability does not significantly change with the degree distribution, decreasing with degree.
    (b) In cooperative communities ($\mu<0$), species with the degree $k_* =  \lambda K/(|\mu|m)$ are most likely to go extinct.
    As $\mu$ decreases ($|\mu|$ increases), $k_*$ becomes smaller until $k_*\rightarrow 0$ at the threshold $\mu=\mu_c$ where $m$ diverges indicating the UG phase.
    We used $\mu = -0.9, -0.6$, and $-0.1$ selected near the threshold $\mu_c \approx -0.968, -0.652$, and $-0.191$ for the Poisson and the two power-law degree distributions, respectively.}
    \label{fig:fig4}
\end{figure}

\begin{figure*}[!ht]
    \centering
    \includegraphics[width=\linewidth]{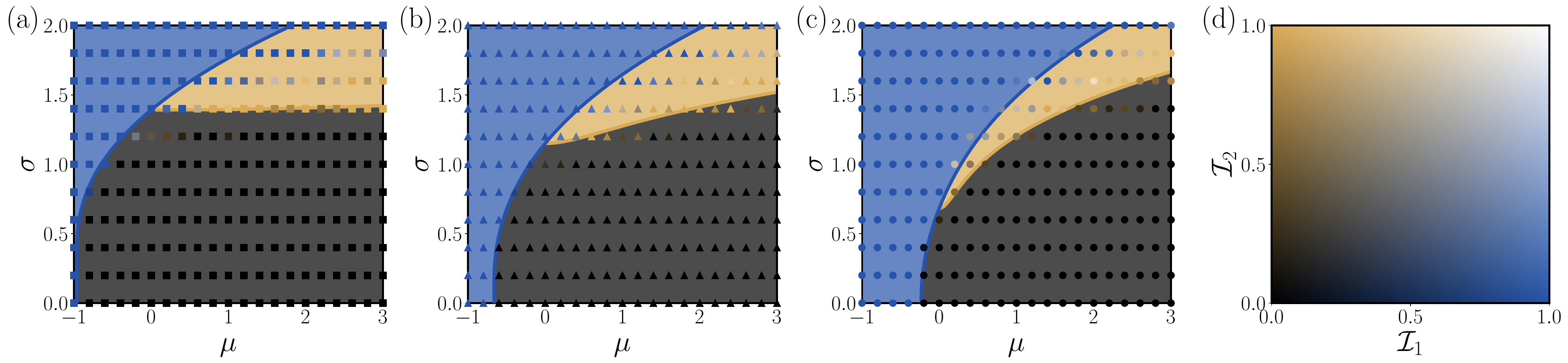}
    \caption{Phase diagrams of the unique fixed point (UFP, black), unbounded growth (UG, blue), and multiple attractors (MA, yellow) phases in the ($\mu, \sigma$) plane for different degree distributions, (a) $P(k) = \text{Pois}(k;K)$, (b) $P(k)\sim k^{-3.5}$, and (c) $P(k)\sim k^{-2.5}$. 
    The background color represents the theoretical prediction and the colors of dots depend on two indices $\mathcal{I}_1$ and $\mathcal{I}_2$ measured in simulations as given in (d). 
    One can see a good agreement between theory and simulations.
    The indices $\mathcal{I}_1$ and $\mathcal{I}_2$ detect the UG and MA phase, respectively (see Sec. II.D. in SM~\cite{supp} for the details).
    Altogether, $\mathcal{I}_1\simeq 1$ with $\mathcal{I}_2\simeq 0$ indicates the UG phase, and $\mathcal{I}_1\simeq 0$ with $\mathcal{I}_2\simeq 1$ indicates the MA phase. 
    If both indices are nearly zero, we identify the phase as the UFP phase.
    }
    \label{fig:fig5}
\end{figure*}

The counterintuitive drop of diversity with cooperation can be intuitively understood in a star-like network, where numerous ``peripheral'' species interact with a small number of ``hub'' species.
The influences of the peripheral species on the hub, when summed up, are likely to be cooperative on average with a small fluctuation due to the central limit theorem.  
In contrast, from the viewpoint of peripherical species, the effect from the hub fluctuates under $J_{ij}\sim\mathcal{N}(\mu/K,\sigma^2/K)$.
Therefore, the hub is very likely to benefit from the peripheral species while hubs could help or suppress the growth of peripheral species by fluctuations.
Moreover, the expectedly large abundance of the hub can threaten a peripheral species even if only a weak competitive interaction is present between them.
This is highly contrasted to the case of no significant heterogeneity in degrees, in which all species, topologically similar, essentially impose an averaged effect on one another with similar abundances.
\newline

{\it Abundance and survival of individual species.---}
Given such difference between hub and peripheral species, we investigate how the fate of individual species is differentiated by their degrees.
Using the HDMFT solution in Eq.~\eqref{eq:CondProb}, we calculate the survival probability of a species with degree $k$, represented by a cumulative Gaussian distribution $\phi(k) = \int_{-\infty}^\infty dx \: \Theta(x) \rho(x|k) = (2\pi)^{-1/2}\int_{-\infty}^{\Delta(k)} dz \exp(-z^2/2)$ with
\begin{align}
    \Delta(k) = \frac{1}{\sigma\sqrt{q}} \left[\lambda\sqrt{\frac{K}{k}} -  \mu m \sqrt{\frac{k}{K}} \right]\,.
    \label{eq:Delta}
\end{align}
The two terms in Eq.~\eqref{eq:Delta} represent the self-growth and the expected influences of $k$ neighboring species, respectively, as seen in rewriting Eq.~\eqref{eq:CondProb} as $x(k) = \sigma \sqrt{qk/K} \max(0,\Delta(k)-z)$.  
Each term dominates for $k\ll k_*$ and $k\gg k_*$, respectively, with the characteristic degree $k_* \equiv \lambda K/(|\mu|m)$. 

For the competitive case with $\mu>0$, species with large $k$ may have negative $\Delta(k)$ and thereby have small $\phi(k)$.
Specifically, the survival probability $\phi(k)$ sharply decreases in the vicinity of $k_*$ [see Fig.~\ref{fig:fig4}(a)].
Furthermore, a strong competition drags $k_*$ down, manifested as decreasing $\phi$ with respect to $\mu$ in Fig.~\ref{fig:fig3}(a).
On the other hand, for the cooperative case with $\mu<0$, $\Delta(k)$ becomes positive for all $k$, resulting in $\phi(k) > 1/2$. 
For the species with as small degree as $k\ll k_*$ or as large as $k\gg k_*$, the self-growth or the expected sum of the partners' influences is likely to dominate the fluctuation of the latter, resulting in large $\Delta(k)$ and $\phi(k)$.  
If a species' degree $k$ is comparable to $k_*$, the fluctuation can be so significant as to make $x(k)$ small or zero, reducing $\Delta(k)$ and $\phi(k)$. 
We also find that due to $k_*$ decreasing with $\mu$, low-degree species are more likely to be eliminated under stronger cooperation [see Fig.~\ref{fig:fig4}(b)].
\newline

{\it Phase diagram.---}
For a comprehensive comparison with previous studies~\cite{bunin2017ecological, galla2018dynamically}, we investigate how the degree heterogeneity affects the phase diagram consisting of the unique fixed point (UFP), unbounded growth (UG), and multiple attractors (MA) phases in the $(\mu,\sigma)-$plane by using both numerical solutions to Eq.~\eqref{eq:LV} and the HDMFT prediction. 
The results in Fig.~\ref{fig:fig5} first reveal that the UFP phase shrinks as the degree heterogeneity increases.
The triple point is given by $(\mu_t, \sigma_t) = (0,\sqrt{2K^2/\langle k^2 \rangle})$ from the HDMFT approach, which converges to $(0,0)$ for $P(k)\sim k^{-\gamma}$ with $2<\gamma<3$ due to the divergence of the second moment $\langle k^2\rangle$ (See Sec. I.F. in SM~\cite{supp}).
Consequently, the system cannot transit directly from the UFP to the UG phase under such strong heterogeneity.  
Note that the triple point in Fig.~\ref{fig:fig5}(c) does not precisely converge to $(0, 0)$ due to finite-size effects.
\newline

{\it Summary and discussion.---}
We have proposed a novel MF framework capable of addressing two types of heterogeneity in the random Lotka-Volterra systems: the interaction strength with $\mathbf{J}$ and the interaction structure with $\mathbf{A}$.
The obtained MF dynamics reveals the differentiation of the individual species' behaviors depending on their degrees, helping us to understand the presence of the characteristic degree at which the randomness of strength is the most dominant and the origin of nontrivial emergent features of the whole community when both types of heterogeneity are present. 

As a final cautionary remark, our approach is based on the moment-generating functional expanded up to the leading order of $K^{-1}$.
The higher-order terms such as $\sim K^{-2}$ and quartic interactions $\sim x^4$ should be considered to study the case of $K$ not large enough, the investigation of which will clarify the effects of interaction sparsity.
Furthermore, it will allow us to investigate whether our results that degree heterogeneity is a prerequisite for causing diversity drops in cooperative communities still hold for even smaller $K$. 

Our framework can be extended in diverse directions, including the incorporation of general distributions of interaction strength, beyond the Gaussian distribution.
Another challenging work can be understanding the topological properties of the community of surviving species, including the size and number distributions of the possibly fragmented subcommunities. Finally, we would like to emphasize the broad applicability of our method beyond natural ecosystems, as long as a system is connected by networks and describable through simple equations like GRLV. This includes diverse domains such as economic or political ecosystems, demonstrating the relevance of our theoretical framework across various fields.

\begin{acknowledgments}
This work was supported by the National Research Foundation (NRF) of Korea under Grant Nos.~NRF-2021R1C1C1004132 (S.H.L.), NRF-2022R1A4A1030660 (S.H.L.), NRF-2019R1A2C1003486 (D.-S.L.), NRF-RS-2023-00214071 (H.J.P.), and a KIAS Individual Grant(No. CG079902) at Korea Institute for Advanced Study (D.-S.L.) This work was supported by INHA UNIVERSITY Research Grant as well (H.J.P.)
\end{acknowledgments}
\end{CJK}
%%%%%%%%%%%%%%%%%%%%%%%%%%%%%%%%%%%%%%%%%%%
%\bibliographystyle{apsrev4-1}
%\bibliography{reference.bib}
%merlin.mbs apsrev4-1.bst 2010-07-25 4.21a (PWD, AO, DPC) hacked
%Control: key (0)
%Control: author (72) initials jnrlst
%Control: editor formatted (1) identically to author
%Control: production of article title (-1) disabled
%Control: page (0) single
%Control: year (1) truncated
%Control: production of eprint (0) enabled
%
%%%%%%%%%%%%%%%%%%%%%%%%%%%%%%%%%%%%%%%%%%%

\end{document}

\title{Incorporating Heterogeneous Interactions for Ecological Biodiversity: Supplemental Material}

\author{Jong Il Park}
\affiliation{Department of Physics, Inha University, Incheon 22212, Korea}
\author{Deok-Sun Lee}\email[Corresponding author: ]{deoksunlee@kias.re.kr}
\affiliation{School of Computational Sciences, Korea Institute for Advanced Study, Seoul 02455, Korea}
\author{Sang Hoon Lee}\email[Corresponding author: ]{lshlj82@gnu.ac.kr}
\affiliation{\mbox{Department of Physics and Research Institute of Natural Science, Gyeongsang National University, Jinju 52828, Korea}}
\affiliation{\mbox{Future Convergence Technology Research Institute, Gyeongsang National University, Jinju 52849, Korea}}
\author{Hye Jin Park}
\email[Corresponding author: ]{hyejin.park@inha.ac.kr}
\affiliation{Department of Physics, Inha University, Incheon 22212, Korea}

\maketitle
\tableofcontents
\clearpage

\section{Heterogeneous Dynamical Mean-Field Theory (HDMFT)}
\subsection{Derivation with moment-generating functional}
For a dynamical system with disorders, the Martin-Siggia-Rose-De Dominicis-Janssen (MSRDJ) formalism or generating functional method is useful to analyze a given system.
By applying this method, we derive the abundance dynamics in $S-$species random Lotka-Volterra system, described as
\begin{align}
    \dot{x}_i = x_i \left( \lambda - x_i - \sum_{j\backslash i}J_{ij}A_{ij}x_j\right) \equiv f_i(\mathbf{x};\lambda,\mathbf{J},\mathbf{A}),
    \label{eq:LV}
\end{align}
where $\mathbf{J}$ and $\mathbf{A}$ represent the interaction strength and adjacency matrices, respectively, and $\sum_{j\backslash i}$ denotes the summation over $j$ except for $j=i$.

We consider the two types of quenched disorders in $J_{ij}$ and $A_{ij}$.
The interaction strength $J_{ij}$ is a Gaussian random variable that follows
\begin{align}
    {\langle J_{ij}\rangle}_\mathbf{J} = \mu/K, \quad {\llangle J_{ij}^2\rrangle}_\mathbf{J} = \sigma^2/K, \quad\mathrm{and}\quad {\llangle J_{ij}J_{ji} \rrangle}_\mathbf{J} &= r\sigma^2/K,
\end{align}
where ${\langle \cdots \rangle}_\mathbf{J}$ and ${\llangle \cdots \rrangle}_{\mathbf{J}}$ denote the moment and cumulant for an interaction ensemble \{$\mathbf{J}_1, \mathbf{J}_2, \cdots$\}, respectively.
We scale the interaction parameters by the mean degree $K = S^{-1} \sum_{i,j} A_{ij}$.
Dividing Eq.~\eqref{eq:LV} by $x_i$ and discretizing time into a sequence of intervals with $\Delta t$ lead to the following path-integral formulation of moment-generating functional:
\begin{equation}
\begin{aligned}
    Z[\boldsymbol{\psi},\boldsymbol{\hat{\psi}}](\mathbf{J},\mathbf{A}) &= \lim_{\Delta t \rightarrow 0}\prod_{t}\int d\mathbf{x}(t)\:e^{\mathbf{x}(t) \cdot \boldsymbol{\psi}(t)\Delta t}\delta\left[\Delta\mathbf{x} - \left(\mathbf{f}(\mathbf{x}(t);\lambda,\mathbf{J},\mathbf{A})+ \hat{\boldsymbol{\psi}}(t)\right)\Delta t\right]\\ 
    &= \int \mathcal{D}[\mathbf{x},i\hat{\mathbf{x}}]\:\exp \left\{\sum_{i} \int dt \left[\hat{x}_i\left( \frac{\dot{x}_i}{x_i}-\lambda+x_i+\sum_{j\backslash i} J_{ij}A_{ij}x_j \right)+\hat{x}_i\hat{\psi}_i+x_i\psi_i\right]\right\},
    \label{Seq:MGF}
\end{aligned}
\end{equation}
where $\int \mathcal{D}[\mathbf{x},i\mathbf{\hat{x}}] = \int_{-\infty}^\infty \int_{-i\infty}^{i\infty}\prod_t d\mathbf{x}(t)\:d\hat{\mathbf{x}}(t)/(2\pi i)$ is a functional integral, and $-\boldsymbol{\hat{\psi}}(t)$ is an external perturbation that evaluates the response function of the system, which will be taken as zero later.
To derive the mean-field (MF) dynamics, we average the moment-generating functional over the interaction strength ensemble $\{\mathbf{J}_1, \mathbf{J}_2, \cdots\}$ and the network structure ensemble \{$\mathbf{A}_1, \mathbf{A}_2, \cdots$\}.
Note that the disorders $J_{ij}$ and $A_{ij}$ are independent, so the order of the averaging process does not matter.

We first perform the averaging over the network ensemble \{$\mathbf{A}_1, \mathbf{A}_2, \cdots$\}.
The adjacency matrix has binary values as $A_{ij}=1$ if two nodes $i$ and $j$ are connected, and $A_{ij}=0$ otherwise.
Let $p_{ij}$ be the probability that two nodes $i$ and $j$ are connected. Then the average of the disorder-dependent term in Eq.~\eqref{Seq:MGF} is given by
\begin{equation}
\begin{aligned}
    \overline{ \exp \left\{ \sum_{i\neq j} \int dt\:\hat{x}_i(t) J_{ij}A_{ij}x_j(t)\right\}} &= \prod_{i<j} \left[(1-p_{ij}) + p_{ij} {\left\langle\exp\left\{\int dt\: 
    \hat{x}_i(t) J_{ij} x_j(t) + \int dt\: \hat{x}_j(t) J_{ji} x_i(t)\right\}\right\rangle}_\mathbf{J}\right]\\
    & = \exp\left\{\sum_{i< j}\log \left[ 1+ p_{ij}{\left\langle\exp Q_{ij} - 1\right\rangle}_\mathbf{J}\right]\right\}\\
    & \approx \exp\left\{\sum_{i< j}p_{ij}{\left\langle\exp Q_{ij} - 1\right\rangle}_\mathbf{J}\right\} \,,
    \label{eq:avgA}
\end{aligned}
\end{equation}
where we introduce $\overline{\cdots}$ for the average over entire disorders, and $Q_{ij}$ is the symmetrized Gaussian random variable
\begin{align}
    Q_{ji} = Q_{ij} \equiv \int dt\: \hat{x}_i(t) J_{ij} x_j(t) + \int dt\: \hat{x}_j(t) J_{ji} x_i(t) \,.
\end{align}
The network ensemble \{$\mathbf{A}_1, \mathbf{A}_2, \cdots$\} we consider is a collection of random networks where each degree sequence $\{k_i\}$ is drawn from a given degree distribution $P(k)$.
In this case, the connection probability is given by $p_{ij} = k_ik_j/SK + O(S^{-2})$.
Plugging this expression into Eq.~\eqref{eq:avgA} and taking the average over the interaction ensemble \{$\mathbf{J}_1, \mathbf{J}_2, \cdots$\} gives
\begin{equation}
\begin{aligned}
    \overline{ \exp \left\{\sum_{i\neq j} \int dt\: \hat{x}_i(t) J_{ij} A_{ij} x_j(t)\right\}} &\approx \exp \left\{\sum_{i< j } \frac{k_ik_j}{SK}\left[\exp\left({\langle Q_{ij} \rangle}_\mathbf
    {J} + \frac{1}{2} {\llangle Q_{ij}^2 \rrangle}_\mathbf{J} \right)-1\right]\right\}\\
    &= \exp \left\{\sum_{i< j} \sum_{n=1}^\infty \frac{k_ik_j}{SK}\left({\langle Q_{ij}\rangle}_\mathbf{J} + \frac{1}{2}{\llangle Q_{ij}^2 \rrangle}_\mathbf{J} \right)^n \right\} \,.
    \label{eq:avg}
\end{aligned}
\end{equation}
The statistical properties ${\langle Q_{ij}\rangle}_\mathbf{J}$ and ${\llangle Q_{ij}^2\rrangle}_\mathbf{J}$ scale as $\mathcal{O}(K^{-1})$ within the range of overall parameters' scales: $\mu$, $\sigma$, $r$ $\sim \mathcal{O}(1)$.
Taking the leading order of $K^{-1}$ gives
\begin{equation}
\begin{aligned}
    &\overline{\exp \left\{\sum_{i\neq j} \int dt\: \hat{x}_i(t) J_{ij} A_{ij} x_j(t)\right\}}\\
    & \qquad = \exp \left\{ \sum_{i\neq j}\frac{k_ik_j}{SK}\left[\frac{\mu}{K}\int dt\: \hat{x}_i(t) x_j(t) + \frac{\sigma^2}{2K}\int dt\:dt'\: \hat{x}_i(t)\hat{x}_i(t')x_j(t)x_j(t')\right. \right.\\
    &\hspace{6.7cm}\left. \left.+ \frac{r\sigma^2}{2K} \int dt\:dt'\: \hat{x}_i(t)x_i(t') x_j(t)\hat{x}_j(t') + \mathcal{O}(K^{-2})\right]\right\} \,.
    \label{Seq:Korder}
\end{aligned}
\end{equation}

By introducing the following macroscopic quantities:
\begin{equation}
\begin{aligned}
    m(t) &= S^{-1}\sum_i \frac{k_i}{K} x_i(t),\\
    w(t) &= S^{-1}\sum_i \frac{k_i}{K} \hat{x}_i(t),\\
    q(t,t') &= S^{-1}\sum_i \frac{k_i}{K} x_i(t)x_i(t'),\\
    g(t,t') &= S^{-1}\sum_i \frac{k_i}{K} x_i(t)\hat{x}_i(t'),\\
    b(t,t') &= S^{-1}\sum_i \frac{k_i}{K} \hat{x}_i(t) \hat{x}_i(t')\,.
\end{aligned}
\end{equation}
we rewrite the disorder-averaged term as
\begin{equation}
\begin{aligned}
    &\overline{\exp \left\{\sum_{i\neq j} \int dt\:\hat{x}_i(t) J_{ij} A_{ij}x_j(t)\right\}}\\
    &\qquad = \exp \left\{S \left[\mu \int dt\:w(t)m(t) + \frac{1}{2}{\sigma}^2 \int dt\:dt'\:\left[b(t,t')q(t,t') + rg(t,t')g(t',t)\right] \right] + R\right\}\,,
\end{aligned}
\end{equation}
where the diagonal contribution $R\sim S^{-1}\sum_i k_i^2<\mathcal{O}(S)$ for finite $K$ so that it can be neglected.
The macroscopic quantities can formally be introduced into the moment-generating functional using the Dirac delta functions in their integral representation, for example,
\begin{equation}
\begin{aligned}
    1 &= \int \mathcal{D}[q]\: \prod_{t,t'} \delta \left[ Sq(t,t') - \sum_i \frac{k_i}{K} x_i(t) x_i(t')\right]\\
    &= \int \mathcal{D}[q,i\hat{q}]\:\exp \left\{\int dt\:dt'\:\hat{q}(t,t')\left[ Sq(t,t') - \sum_i \frac{k_i}{K} x_i(t)x_i(t')\right]\right\}\,.
\end{aligned}
\end{equation}
The disorder-averaged functional can be rewritten in macroscopic quantities and their conjugates, 
\begin{equation}
\begin{aligned}
    \overline{Z[\boldsymbol{\psi},\boldsymbol{\hat{\psi}}]} &\equiv \int \mathcal{D}[\boldsymbol{\theta},i\boldsymbol{\hat{\theta}}]\:\exp\Gamma[\boldsymbol{\theta},\boldsymbol{\hat{\theta}}]\\ 
    &= \int \mathcal{D}[\boldsymbol{\theta},i\boldsymbol{\hat{\theta}}]\:\exp\left(\Phi[\boldsymbol{\theta},\boldsymbol{\hat{\theta}}] + \Psi[\boldsymbol{\theta}] + \log Z_0[\boldsymbol{\hat{\theta}}]\right)\,,
    \label{Seq:Zbar}
\end{aligned}
\end{equation}
where $\boldsymbol{\theta} \equiv (m,w,q,g,b)$, $\boldsymbol{\hat{\theta}} \equiv (\hat{m},\hat{w},\hat{q},\hat{g},\hat{b})$, and the corresponding potentials
\begin{equation}
\begin{aligned}
    &\Phi = S\left\{\int dt\:\left[\hat{m}(t)m(t) + \hat{w}(t)w(t)\right]\right.\\ 
    &\hspace{2.7cm} + \left.\int dt\:dt'\:\left[\hat{q}(t,t')q(t,t')+\hat{g}(t,t')g(t,t')+\hat{b}(t,t')b(t,t')\right]\right\}\,,
\end{aligned}
\end{equation}
\begin{equation}
\begin{aligned}
    \Psi = S\left\{\mu\int dt\:w(t)m(t) + \frac{1}{2}\sigma^2 \int dt\:dt'~\left[b(t,t')q(t,t')+rg(t,t')g(t',t)\right]\right\}\,,
\end{aligned}
\end{equation}
and
\begin{equation}
\begin{aligned}
    Z_0[\boldsymbol{\hat{\theta}}] &= \int \mathcal{D}[\mathbf{x},i\mathbf{\hat{x}}]\: \exp\left\{\sum_i \int dt\:\left[\hat{x}_i(t) \left(\frac{\dot{x}_i(t)}{x_i(t)} - \lambda + x_i(t)\right)\right] + \sum_i \int dt\:[x_i(t)\psi_i(t) + \hat{x}_i(t) \hat{\psi}_i(t)]\right\}\\
    &\hspace{1.8cm}\times\exp\left\{- \sum_i \frac{k_i}{K}\int dt\:\left[\hat{m}(t)x_i(t) + \hat{w}(t)\hat{x}_i(t)\right]\right\}\\
    &\hspace{1.8cm}\times\exp\left\{-\sum_i\frac{k_i}{K}\int dt\:dt'\:\left[\hat{q}(t,t')x_i(t)x_i(t') + \hat{g}(t,t')x_i(t)\hat{x}_i(t') + \hat{b}(t,t')\hat{x}_i(t)\hat{x}_i(t')\right]\right\}\\
    &\equiv\int \mathcal{D}[\mathbf{x},i{\mathbf{\hat{x}}}]\: \exp\left\{\Gamma_0[\mathbf{x},\mathbf{\hat{x}}] + \mathbf{x}\cdot\boldsymbol{\psi} + \mathbf{\hat{x}}\cdot \boldsymbol{\hat{\psi}}\right\}\,.
\end{aligned}
\end{equation}

In the thermodynamic limit $S\rightarrow\infty$, the functional integral in Eq.~\eqref{Seq:Zbar} is dominated by the saddle-point value, i.e., $\overline{Z[\boldsymbol{\psi},\boldsymbol{\hat{\psi}}]} \approx \exp\Gamma[\boldsymbol{\theta}^*,\boldsymbol{\hat{\theta}}^*]$, where $\left.\nabla\right|_{\boldsymbol{\theta}^*,\boldsymbol{\hat{\theta}}^*}\Gamma[\boldsymbol{\theta},\boldsymbol{\hat{\theta}}] = 0$.
The saddle-point condition with respect to $\boldsymbol{\theta}$ gives
\begin{equation}
\begin{aligned}
    \frac{\delta \Gamma}{\delta m} = 0 &\Longleftrightarrow \hat{m}(t)+\mu w(t) = 0,\\
    \frac{\delta \Gamma}{\delta w} = 0 &\Longleftrightarrow \hat{w}(t)+\mu m(t) = 0,\\
    \frac{\delta \Gamma}{\delta q} = 0 &\Longleftrightarrow \hat{q}(t,t')+\frac{1}{2}\sigma^2 b(t,t') = 0,\\
    \frac{\delta \Gamma}{\delta g} = 0 &\Longleftrightarrow \hat{g}(t,t')+r\sigma^2 g(t',t) = 0,\\
    \frac{\delta \Gamma}{\delta b} = 0 &\Longleftrightarrow \hat{b}(t,t')+\frac{1}{2}\sigma^2 q(t,t') = 0\,.
\end{aligned}
\end{equation}
Next, the saddle-point condition with respect to $\boldsymbol{\hat{\theta}}$ gives 
\begin{equation}
\begin{aligned}
    \frac{\delta \Gamma}{\delta \hat{m}} = 0 &\Longleftrightarrow m(t)- S^{-1}\sum_i \frac{k_i}{K} {\langle x_i(t) \rangle}_0 = 0,\\
    \frac{\delta \Gamma}{\delta \hat{w}} = 0 &\Longleftrightarrow w(t)-S^{-1}\sum_i \frac{k_i}{K} {\langle \hat{x}_i(t) \rangle}_0 = 0,\\
    \frac{\delta \Gamma}{\delta \hat{q}} = 0 &\Longleftrightarrow q(t,t')-  S^{-1}\sum_i \frac{k_i}{K} {\langle x_i(t)x_i(t') \rangle}_0 = 0,\\
    \frac{\delta \Gamma}{\delta \hat{g}} = 0 &\Longleftrightarrow g(t,t')-S^{-1}\sum_i \frac{k_i}{K} {\langle x_i(t) \hat{x}_i(t') \rangle}_0 = 0,\\
    \frac{\delta \Gamma}{\delta \hat{b}} = 0 &\Longleftrightarrow b(t,t')-S^{-1}\sum_i \frac{k_i}{K} {\langle \hat{x}_i(t) \hat{x}_i(t') \rangle}_0 = 0\,,
\end{aligned}
\end{equation}
where ${\langle \cdots \rangle}_0\equiv \int \mathcal{D}[\mathbf{x},i{\mathbf{\hat{x}}}]\:(\cdots)\exp\Gamma_0[\mathbf{x},\mathbf{\hat{x}}]/ \int \mathcal{D}[\mathbf{x},i\mathbf{\hat{x}}]\:\exp \Gamma_0[\mathbf{x},\mathbf{\hat{x}}]$.
Due to the normalization condition, the moment-generating functional for a particular path $Z[\boldsymbol{\psi}=0,\boldsymbol{\hat{\psi}}]=1$.
This makes some of the macroscopic quantities vanish:
\begin{equation}
\begin{aligned}
    w(t) &= S^{-1}\sum_i \frac{k_i}{K} {\langle \hat{x}_i(t) \rangle}_0 = S^{-1}\sum_{i}\frac{k_i}{K} \left.\frac{\delta \overline{Z[0,\boldsymbol{\hat{\psi}}]}}{\delta \hat{\psi}_i(t)}\right|_{\boldsymbol{\hat{\psi}}=0} = 0,\\
    b(t,t') &= S^{-1}\sum_i \frac{k_i}{K} {\langle \hat{x}_i(t) \hat{x}_i(t') \rangle}_0 = S^{-1}\sum_{i}\frac{k_i}{K} \left.\frac{\delta^2 \overline{ Z[0,\boldsymbol{\hat{\psi}}]}}{\delta \hat{\psi}_i(t)\:\delta \hat{\psi}_i(t')}\right|_{\boldsymbol{\hat{\psi}}=0} = 0\,.
\end{aligned}
\end{equation}
Moreover, the response function $g(t,t') = \partial x(t) / \partial \hat{\psi}(t')$ is zero for $t<t'$ due to causality.
Taking the normalization and causality conditions into account, $\Phi= 0 = \Psi$ at $(\boldsymbol{\theta}^*,\boldsymbol{\hat{\theta}}^*)$, which leads to the effective moment-generating functional 
\begin{equation}
\begin{aligned}
    \overline{Z[\boldsymbol{\psi},\boldsymbol{\hat{\psi}}]} &\approx Z_0[\boldsymbol{\hat{\theta}}^*]\\
    &= \int \mathcal{D}[\mathbf{x},i\mathbf{\hat{x}}]\:\exp\left\{\sum_i \int dt\:\left[\hat{x}_i(t)\left(\frac{\dot{x}_i(t)}{x_i(t)} - \lambda + x_i(t)+\mu\frac{k_i}{K}m(t)\right) + \hat{x}_i\hat{\psi}_i+x_i\psi_i \right]\right\}\\
    &\qquad\times \exp \left\{\sum_i\int dt\:dt'\:\left[\frac{1}{2}\sigma^2 q(t,t')\frac{k_i}{K}\hat{x}_i(t)\hat{x}_i(t')+r\sigma^2 g(t',t)\frac{k_i}{K}x_i(t)\hat{x}_i(t')\right]\right\}\,,
\end{aligned}
\end{equation}
and the corresponding MF dynamics is
\begin{equation}
\begin{aligned}
    \dot{x}_i(t) = x_i(t)\left[\lambda-x_i(t) - \mu \frac{k_i}{K}m(t) -  \sigma\sqrt{\frac{k_i}{K}}\eta_i(t) - r\sigma^2 \frac{k_i}{K} \int dt'\:g(t,t')x_i(t')\right]\,,
\end{aligned}
\end{equation}
or equivalently,
\begin{equation}
\begin{aligned}
    \dot{x}(t;k) = x(t;k)\left[\lambda-x(t;k) - \mu \frac{k}{K}m(t) -  \sigma\sqrt{\frac{k}{K}}\eta(t) - r\sigma^2 \frac{k}{K} \int dt'\:g(t,t')x(t';k)\right]\,,
\end{aligned}
\label{Seq:DMFT}
\end{equation}
where $\eta(t)$ is the zero-mean Gaussian noise, and the macroscopic quantities are
\begin{equation}
\begin{aligned}
    m(t) &= \sum_k P(k) \frac{k}{K}{\langle x(t;k) \rangle}_0,\\
    {\langle \eta(t) \eta(t') \rangle}_0 &= q(t,t') = \sum_k P(k)\frac{k}{K} {\langle x(t;k) x(t';k) \rangle}_0,\\
    g(t,t') &= \sum_k P(k) \frac{k}{K} {\left\langle\frac{\partial x(t;k)}{\partial\hat{\psi}(t';k)}\right\rangle}_0\,.
\end{aligned}
\end{equation}

\subsection{Stationary solution}
Now we consider the stationary solution of the MF dynamics.
At the fixed point, $\dot{x}_i=0$, and $\eta_i$ becomes a static Gaussian random variable.
Thus, the abundance at the fixed point is
\begin{align}
    x(k) = 0 \qquad \mathrm{or} \qquad x(k) = \frac{\lambda - \mu m k/K- \sigma\sqrt{q k/K}z}{1-r\sigma \chi k/K}\,,
\end{align}
where the static variables $m$, $q$, and $\chi$ are measured at the stationary state: $ m = \lim_{t\rightarrow\infty} m(t)$, $q = \lim_{t\rightarrow\infty} q(t,t)$ and the integrated response $\chi = -\int_0^\infty d\tau\:g(\tau)$.
When $1-r\sigma^2\chi k/K >0$, the fixed point with the greater value than the other becomes stable (linear stability will be checked in Sec.~\ref{sec:MA_phase}.), which leads to
\begin{align}
    x(k) = \max\left(0, \frac{\lambda - \mu m k/K -\sigma \sqrt{q k/K}z}{1-r\sigma^2 \chi k/K} \right).
    \label{Seq:fp}
\end{align}
The result indicates that for each species with a given degree $k$, its abundance follows the truncated Gaussian distribution, $x(k)\sim\rho(x|k)$.
The MF solution is valid for the unique fixed point (UFP) phase, where the system always converges to a single equilibrium.
Note that the fixed point in Eq.~(\ref{Seq:fp}) does not depend on the initial state.

In the thermodynamic limit, the overall average is equivalent to
\begin{align}
    \langle \cdots \rangle \equiv S^{-1} \sum_i {\langle \cdots \rangle}_0 \longmapsto \sum_k P(k)\int dx\:(\cdots)\:\rho(x|k)\,.
\end{align}
The three macroscopic quantities can be obtained by solving the following self-consistent (SC) equations:
\begin{align}
    \label{Seq:SCm}
    m &= \langle kx\rangle/K = \sigma\sqrt{q} \left[\sum_k P(k) \left(\frac{k}{K}\right)^{3/2} \int_{-\Delta(k)}^\infty Dz\: \frac{z+\Delta(k)}{1-r\sigma^2 \chi k/K}\right]\,,\\
    \label{Seq:SCq}
    q &= \langle kx^2 \rangle/K = \sigma^2 q \left[\sum_k P(k)\left(\frac{k}{K}\right)^2 \int_{-\Delta(k)}^\infty Dz\: \left(\frac{z+\Delta(k)}{1-r\sigma^2 \chi k/K}\right)^2\right]\,,\\
    \label{Seq:SCchi}
    \chi &= -\left\langle k x \hat{x}\right\rangle/ K = \sum_k P(k) \frac{k}{K} \int_{-\Delta(k)}^\infty Dz\:\frac{1}{1-r\sigma^2 \chi k/K}\,,
\end{align}
where $Dz = dz\:e^{-z^2/2}/\sqrt{2\pi}$ is a Gaussian measure and $\Delta(k) = (\lambda - \mu m k/K)/[\sigma\sqrt{q k/K}]$.
Another major quantity is the average survival probability
\begin{align}
    \phi = \langle \Theta(x) \rangle = \sum_k P(k) \int^\infty_{-\Delta(k)} Dz\,.
\end{align}
From now on, calculations and simulations are taken for the $r=0$ case unless we explicitly mention the reciprocity $r$, where $\chi$ no longer contributes to the SC solutions of $m$ and $q$.

\subsection{Survival probability on regular random networks}
\begin{figure}[!h]
    \centering
    \includegraphics[width=0.5\linewidth]{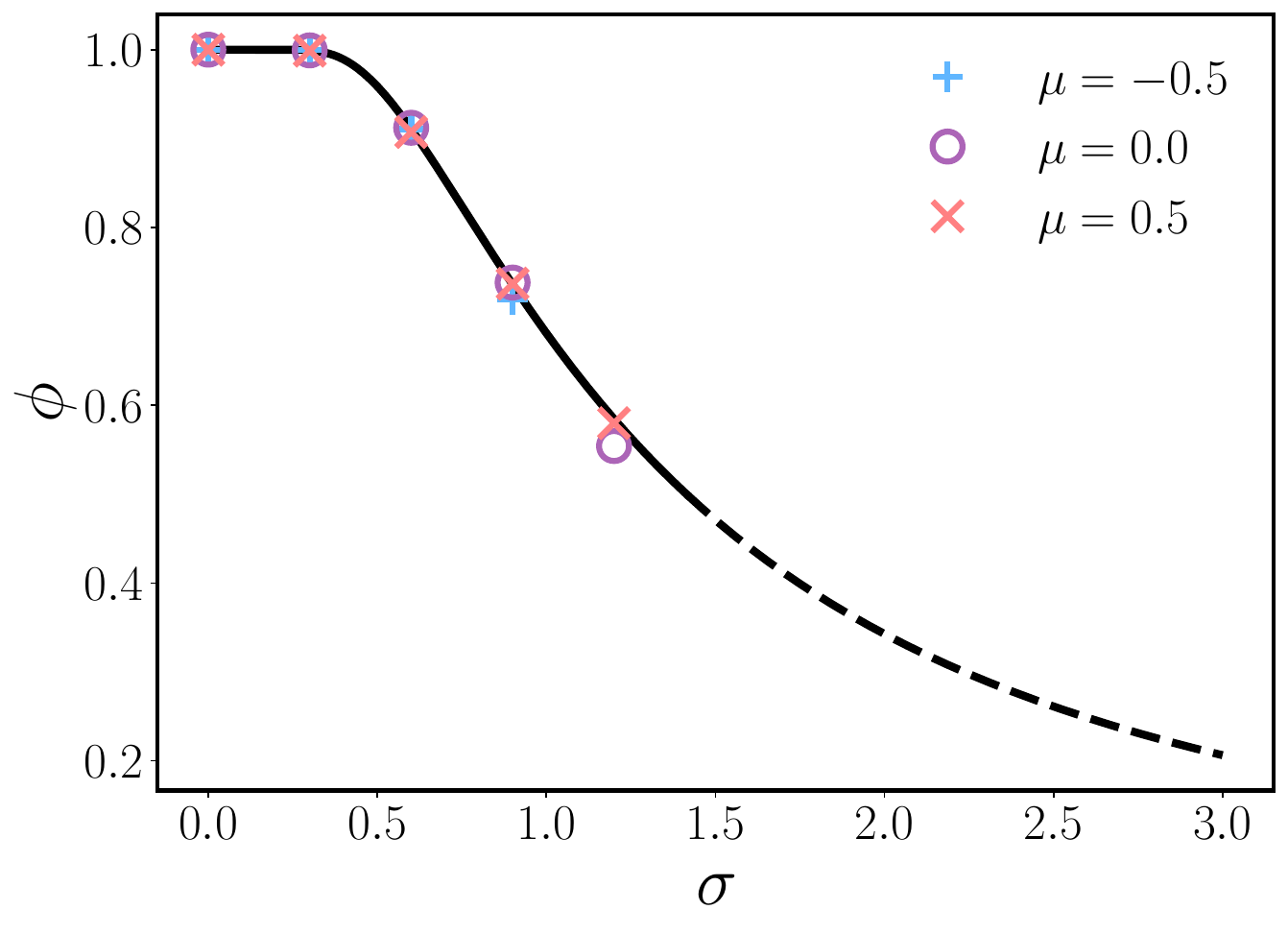}
    \caption{The MF solution and simulation results of $\phi$ with the regular random structure $P(k) = \delta_{k,K}$. The curve represents the MF solution at the unique fixed point phase (solid) and the multiple attractors phase (dashed). The simulation results with various values of $\mu = -0.5$, $0$, and $0.5$ collapse onto a single MF curve, supporting the fact that the survival probability on regular random network structure does not depend on $\mu$.}
    \label{fig:S1}
\end{figure}

From Eq.~\eqref{Seq:SCq} with the regular random network structure $P(k) = \delta_{k,K}$, we obtain a simple equation for $\Delta = (\lambda - \mu m)/[\sigma \sqrt{q}]$ as
\begin{align}
    W(\Delta) \equiv \int_{-\Delta}^\infty Dz\: {(z+\Delta)}^2 = \sigma^{-2}\,,
\end{align}
which leads to
\begin{align}
    \frac{dW(\Delta)}{d\Delta} = 2\int_{-\Delta}^\infty Dz\:(z+\Delta) = 2\langle x\rangle > 0\,.
\end{align}
Here, $W$ is a positive-definite monotonically increasing function, which always guarantees a unique solution $\Delta = W^{-1}(\sigma^{-2})$.
Therefore, the average survival probability $\phi = \int_{-W^{-1}(\sigma^{-2})}^\infty Dz$ is determined by the value of $\sigma$ regardless of $\mu$ values.
The numerical simulations of $\phi$ in Fig.~\ref{fig:S1} with different values of $\mu$ follow the MF prediction.
The MF equations on the regular random networks are identical to those on the fully-connected structures, and thus this argument is consistent with the result reported in the previous study~\cite{bunin2017ecological}. 

\subsection{Transition to the unbounded growth (UG) phase}
\label{sec:UG_phase}
The average abundance $\langle x \rangle$ or its weighted version $m = \langle kx\rangle/K$ starts to diverge when the average cooperation level in the community exceeds a threshold value. 
We first consider the case when the interaction strength is completely uniform, i.e., $\sigma = 0$.
Then the abundance at the stationary state is given by $x(k) = \max(0,\lambda - \mu m k/K)$.
For $\mu<0$, the nonzero fixed point is always positive, and thus the feasible fixed point is always stable, allowing $x(k) = \lambda - \mu m k/K$.
Multiplying $k/K$ and taking the overall average at both sides gives $m = \langle kx\rangle/K = \lambda - \mu m \langle k^2\rangle /K^2$.
Therefore, 
\begin{align}
    m = \frac{\lambda}{1+\mu\langle k^2 \rangle /K^2} = \lambda {\left(\frac{\mu_c - \mu}{\mu_c}\right)}^{-1}~~\text{for}~~\mu<0\,.
\end{align}
Notice that $m$ has a singularity at $\mu_c = - K^2 / \langle k^2 \rangle$.
Consequently, the average abundance also diverges at the singularity of $m$ because $\langle x \rangle = \lambda + |\mu| m$ for $\mu < 0$.

For the case of heterogeneous interaction strength with $\sigma \neq 0$, it becomes complicated to find the exact singularity.
In this case, let us rewrite Eq.~\eqref{Seq:DMFT} with the reduced abundance $\tilde{x}_i = x_i/m$, which makes the weighted average be unity, as $S^{-1} \sum_i (k_i/K)\tilde{x}_i = 1$.
Thus, at the stationary state $\dot{x}_i = 0$,
\begin{align}
    \tilde{x}(k) = \max\left(0, \tilde{\lambda} - \mu\frac{k}{K} - \sigma \sqrt{\tilde{q} \frac{k}{K}} z \right)\,,
    \label{Seq:xtilde}
\end{align}
where the reduced parameters are $\tilde{\lambda} = \lambda/m$ and $\tilde{q} = q/{m}^2 = \langle k \tilde{x}^2 \rangle/K$.
At the unbounded growth (UG) phase, the weighted-average abundance $m$ diverges, equivalent to $\tilde{\lambda} = 0$.
Therefore, the UG phase boundary satisfies the following two identities:
\begin{align}
    \label{Seq:ug1}
    1 &= \frac{\langle k\tilde{x}\rangle}{K} = \sigma\sqrt{\tilde{q}}\left[\sum_k P(k) \left(\frac{k}{K}\right)^{3/2} \int_{-\Delta(k)}^\infty Dz\: (z + \Delta(k))\right]\,,\\
    \label{Seq:ug2}
    1 &= \frac{\langle k\tilde{x}^2\rangle}{K\tilde{q}} = \sigma^2\left[\sum_k P(k) \left(\frac{k}{K}\right)^2\int_{-\Delta(k)} ^\infty Dz\:{(z+\Delta(k))}^2\right]\,.
\end{align}
These SC equations with reduced abundance are not divergent, so the UG phase boundary can be obtained from the numerical solution of Eqs.~\eqref{Seq:ug1} and~\eqref{Seq:ug2}.

\subsection{Transition to the multiple attractors (MA) phase}
\label{sec:MA_phase}
Now we check the linear stability of the fixed point to detect the MA phase boundary. 
Near the fixed point in Eq.~(\ref{Seq:fp}), let us apply the infinitesimal perturbations as $x_i(t) = x_i^* + \epsilon\,\delta x_i (t)$, $\eta_i(t) = \eta_i^* + \epsilon\,\delta\eta_i(t)$, and $\hat{\psi}_i(t) = \epsilon h_i(t)$, where $\eta_i^* = \sqrt{q}z$.
The linear perturbations satisfy ${\langle \delta x_i(t) \rangle}_0 = 0$, ${\langle \delta \eta_i (t) \rangle}_0 = 0$, ${\langle h_i(t) \rangle}_0 = 0$, and ${\langle h_i(t) h_i(t')\rangle}_0 = \delta(t-t')$.
Near $x_i^\star=0$, collecting the terms of order $\epsilon$ gives the following dynamics of the perturbation:
\begin{align}
    \dot{\delta x}_i(t) = \delta x_i(t) \left[ \lambda - \mu m \frac{k_i}{K} - \sigma\sqrt{q \frac{k_i}{K}}z\right]\,.
\end{align}
The linear perturbation $\delta x_i$ exponentially decreases when $\lambda - \mu m k_i/K - \sigma\sqrt{q k_i/K}z < 0$, which means that the solution $x_i=0$ becomes stable.
This condition coincides with one we presumed in calculating the static SC variables.

Near $x_i^\star \neq 0$, the linear perturbation evolves as
\begin{align}
    \dot{\delta x}_i(t) = x_i^* \left[-\delta x_i(t) -\sigma\sqrt{\frac{k_i}{K}}\,\delta\eta_i(t)-r\sigma^2\frac{k_i}{K}\int dt'\:g(t,t')\,\delta x_i(t') - h_i(t)\right]\,.
    \label{Seq:ptb}
\end{align}
Dividing both sides by $x_i^*$ and writing Eq.~(\ref{Seq:ptb}) in Fourier space gives
\begin{align}
    \left[1 + r \sigma^2\frac{k_i}{K} g(\omega) + \frac{i\omega}{x_i^*} \right]\,\delta x_i(\omega) = -\sigma\sqrt{\frac{k_i}{K}}\,\delta \eta_i(\omega)-h_i(\omega)\,.
    \label{eq:Fourier}
\end{align}
Note that $g(\omega=0) = \int_0^\infty d\tau\:g(\tau) = -\chi$ and
\begin{equation}
\begin{aligned}
    q(t,t') &= {\langle \eta_i(t) \eta_i(t')\rangle}_0 \equiv q + \epsilon^2\,\delta q(t,t')\\
    &= {\langle (\eta_i^*+\epsilon\,\delta\eta_i(t))(\eta_i^*+\epsilon\,\delta\eta_i(t'))\rangle}_0 = q + \epsilon^2{\langle \delta\eta_i(t)\,\delta\eta_i(t')\rangle}_0\\
    &= S^{-1}\sum_i \frac{k_i}{K}{\langle(x_i^*+\epsilon \,\delta x_i(t))(x_i^*+\epsilon\,\delta x_i(t')) \rangle}_0 = q + \epsilon^2 \langle k\,\delta x(t)\, \delta x(t')\rangle/K\,,
\end{aligned}
\end{equation}
which leads to ${\langle |\delta \eta_i(\omega)|^2\rangle}_0 = \langle k\, |\delta x(\omega)|^2 \rangle/K$.
With these identities, at $\omega=0$, multiplying the complex conjugates in Eq.~\eqref{eq:Fourier} and taking the average over quenched disorders leads to
\begin{equation}
\begin{aligned}
    \left[1-r\sigma^2 \frac{k_i}{K} \chi\right]^2 {\langle{|\delta x_i(\omega=0)|}^2\rangle}_0 &= \sigma^2 \frac{k_i}{K} {\langle {|\delta \eta_i(\omega=0)|}^2\rangle}_0 + {\langle{|h_i(\omega=0)|}^2\rangle}_0\\
    &= \sigma^2 \frac{k_i}{K^2} \langle k\, {|\delta x(\omega=0)|}^2 \rangle + 1\,.
\end{aligned}
\end{equation}

Let us assume that a stationary state where the autocorrelation of linear perturbation is time-translational invariant, i.e., ${\langle \delta x_i(t)\, \delta x_i(t') \rangle}_0$ only depends on time difference $\tau = t'-t$. 
The integrated autocorrelation, proportional to the relaxation time, is given by
\begin{align}
{\langle {|\delta x_i(\omega=0)|}^2 \rangle}_0 = 
\begin{cases}
\displaystyle 0 &\text{for } x_i^* = 0\\
\\
\displaystyle \left[ 1 - r\sigma^2 \frac{k_i}{K} \chi \right]^{-2}\left[\sigma^2\frac{k_i}{K^2} \langle k\,{|\delta x(\omega=0)|}^2\rangle + 1\right] \quad &\text{for } x_i^* > 0
\end{cases}.
\end{align}
Expanding Eq.~(\ref{Seq:SCchi}) at $r = 0$, we obtain
\begin{align}
    \chi = \frac{\langle k\Theta(x)\rangle}{K}\left[1 + r\sigma^2 \frac{\langle k^2 \Theta(x)\rangle}{K^2} + \mathcal{O}(r^2)\right].
\end{align}
Finally, we get the weighted average relaxation time for small $r$:
\begin{equation}
\begin{aligned}
    \int_{-\infty}^{\infty} d\tau\:\delta q(\tau) &= \frac{\langle k\,{|\delta x(\omega=0)|}^2 \rangle}{K}\\
    & \hspace{-1cm}= \left[1-\sigma^2\frac{\langle k^2\Theta(x) \rangle}{K^2} - 2r\sigma^4 \frac{\langle k\Theta(x)\rangle\langle k^3\Theta(x)\rangle}{K^4} + \mathcal{O}(r^2) \right]^{-1} \left[1 + 2r\sigma^2 \frac{\langle k^2\Theta(x)\rangle}{K^2} +\mathcal{O}(r^2)\right]\frac{\langle k\Theta(x)\rangle}{K}\,.
    \label{Seq:MA}
\end{aligned}
\end{equation}
Notice that it has a singularity at 
\begin{align}
    \sigma^2_g = \frac{K^2}{\langle k^2\Theta(x)\rangle} \left[1-2r \frac{\langle k \Theta(x)\rangle\langle k^3\Theta(x)\rangle}{{\langle k^2\Theta(x) \rangle}^2} + \mathcal{O}(r^2)\right],    
\end{align}
which means that the fluctuations never decay, so the system does not converge back to the initial stationary state.
This indicates that the system could have multiple fixed points or show chaotic behavior.
The critical value $\sigma_g^2 = 2[1-2r+\mathcal{O}(r^2)]$ on fully-connected (or regular random) networks~\cite{bunin2017ecological,galla2018dynamically} is restored by inserting $\langle k^n \Theta(x)\rangle = K^n\phi$ with $\phi = 1/2$ for $n\in\{1,2,3\}$.
With degree heterogeneity, the phase boundary of the multiple attractors (MA) phase depends on the second moment of degrees of survived species for $r=0$.
With nonvanishing reciprocity, the boundary also depends on the higher-order moments.

\subsection{Triple point}
We can also deduce the location of the triple point.
We assume that the triple point exists at $\mu = 0$.
Due to the UG phase boundary condition, the reduced abundance becomes $\tilde{x}^*(k) = \max(0,\sigma\sqrt{\tilde{q} k/K} z)$.
Consequently, only half of the species survive regardless of their degrees, i.e., $\phi(k)=1/2$.
The triple point should also obey the criterion for the MA phase boundary, which leads to $\sigma = \sqrt{K^2/\langle k^2 \Theta(x)\rangle} = \sqrt{2K^2/\langle k^2 \rangle}$.
The validation of this triple point is completed by checking the obtained point $(\mu_t,\sigma_t) = (0,\sqrt{2 K^2/\langle k^2 \rangle})$ satisfies Eq.~\eqref{Seq:ug2}:
\begin{align}
    \frac{\langle k\tilde{x}^2\rangle}{K\tilde{q}} = \frac{2K^2}{\langle k^2 \rangle} \left[\sum_k P(k) \left(\frac{k}{K}\right)^2 \int_0^\infty Dz\:z^2 \right] = \frac{2K^2}{\langle k^2 \rangle} \frac{\langle k^2\rangle}{2K^2} = 1\,.
\end{align}
Note that $(\mu_t, \sigma_t) = (0,0)$ for power-law degree distribution $P(k) \sim k^{-\gamma}$ with exponent $2<\gamma <3$ due to diverging the second moment of degree $\langle k^2 \rangle$.
The strong degree heterogeneity endows the instability of the MF fixed point, manifested as a decreasing area of the UFP phase in the $(\mu,\sigma)-$plane shown in Fig.~5 (the main text).

\subsection{Surpvival probability drop in $\mu$ depending on degree exponents}
\label{sec:phic}
As $\mu$ decreases from zero, the survival probability $\phi$ may change, exhibiting different behaviors depending on the degree exponents. 
Here, under the small $\sigma$ assumption, we show that the survival probability sharply declines when the exponent falls below a threshold value.
To see this, we calculate the survival probability at $mu=0$ and $\mu_c$, the UFP-UG transition point.

We first focus on $\sigma=0$ case where $\mu_c(\sigma =0) = -K^2/\langle k^2 \rangle$.
Using the results from Sec.~\ref{sec:UG_phase}, the reduced abundance $\tilde{x}(k) = -\mu_c(0) k/K$, and thus its weighted second moment is given by
\begin{align}
    \tilde{q}_c (\sigma = 0) = {[\mu_c(0)]}^2 \frac{\langle k^3\rangle}{K^3} = \frac{K \langle k^3\rangle}{{\langle k^2\rangle}^2}.
\end{align}
For power-law degree distribution $P(k) \sim k^{-\gamma}$, $\tilde{q}_c(0)$ diverges when $2<\gamma<4$.
Nevertheless the singularity in $\tilde{q}_c$, the survival probability $\phi$ at $\mu_c$ is unity when $\sigma=0$.
However, with non-zero $\sigma$, the singularity affects the survival probability at $\mu_c$.
In the following paragraphs, we calculate the survival probability at $\mu=0$ and $\mu=\mu_c$ for small $\sigma$ to see how the diversity changes in $\mu$ for different degree exponents.

We expand reduced parameters for small $\sigma$, $\mu_c(\sigma) = \mu_c(0) + \mu_c^{(1)}(0)\sigma +\cdots$ and $\tilde{q}_c(\sigma) = \tilde{q}_c(0) + \tilde{q}_c^{(1)}(0)\sigma +\cdots$, yielding
\begin{equation}
\begin{aligned}
    \Delta_c(k;\sigma) = -\frac{\mu_c(\sigma)}{\sigma\sqrt{\tilde{q}_c(\sigma)}}\sqrt{\frac{k}{K}} = \left[-\frac{\mu_c(0)}{\sigma\sqrt{\tilde{q}_c(0)}} - \frac{1}{\sqrt{\tilde{q}_c(0)}}\left(\mu_c^{(1)}(0) - \frac{\mu_c(0)\tilde{q}_c^{(1)}(0)}{\tilde{q}_c(0)}\right) + \mathcal{O}(\sigma)\right] \sqrt{\frac{k}{K}} \approx \frac{\sigma_\xi}{\sigma}\sqrt{\frac{k}{K}}
\end{aligned}    
\end{equation}
with the characteristic standard deviation $\sigma_\xi = \sqrt{K^3/\langle k^3 \rangle}$.
The subscripts $0$ and $c$ denote $\mu=0$ and $\mu=\mu_c$, respectively.
Then, the survival probability at $\mu_c$ is written as 
\begin{align}
    \phi_c = \sum_k P(k) \int_{-\Delta_c(k)}^\infty Dz \approx 1 - \frac{1}{2}\sum_k P(k) \sqrt{\frac{2}{\pi} {\left(\frac{\sigma}{\sigma_\xi}\right)}^{2}\frac{K}{k}}\exp\left[-\frac{1}{2}{\left(\frac{\sigma}{\sigma_\xi}\right)}^{-2}\frac{k}{K}\right].
    \label{Seq:phic}
\end{align} 
Carrying the same procedure out at $\mu=0$, one can obtain the survival probability for small $\sigma$
\begin{align}
    \phi_0 = \sum_k P(k) \int_{-\Delta_0(k)}^\infty Dz \approx 1 - \frac{1}{2}\sum_k P(k) \sqrt{\frac{2}{\pi} \sigma^{2}\frac{k}{K}}\exp\left[-\frac{1}{2}\sigma^{-2}\frac{K}{k}\right],
    \label{Seq:phi0}
\end{align}
where we used $\tilde{\lambda} = \tilde{\lambda}_0(0) = 1$, $\tilde{q} = \tilde{q}_0(0) = 1$, and thus $\Delta_0(k;\sigma) \approx (\sqrt{K/k})\tilde{\lambda}_0(0)/[\sigma \tilde{q}_0(0)] = \sigma^{-1} \sqrt{K/k}$.
\begin{figure}[!h]
    \centering
    \includegraphics[width=\textwidth]{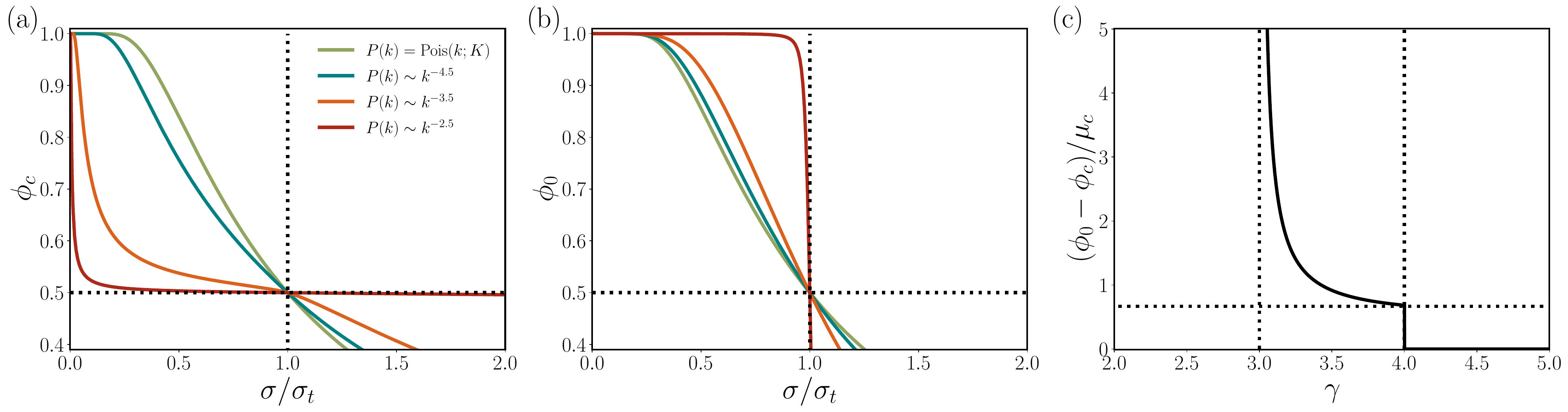}
    \caption{MF solutions of (a) $\phi_c$ and (b) $\phi_0$ with $S = 10^6$ as a function of scaled standard deviation $\sigma/\sigma_t$. The dotted lines denote $\phi = 1/2$ at the triple point $\sigma = \sigma_t$. The survival probability at $\mu_c$ $\phi_c$ rapidly decays to $1/2$ for power-law degree distribution $P(k)\sim k^{-\gamma}$ with $\gamma = 2.5$ and $3.5$ as expected in our approximation. A large plateau of $\phi_0 \approx 1$ appears for $\gamma = 2.5$ in (b). 
    The average slope of survival probability over $[\mu_c,0]$ is drawn in (c). We evaluate the MF solution at $\sigma = 0.001$ in the limit of $S\rightarrow\infty$ using an analytic continuation of $P(k)$. There exists a discontinuous jump from $0$ to $\langle k^2\rangle /(2K^2) = 2/3$ at $\gamma=4$, marked as a horizontal dotted line.}
    \label{fig:S2}
\end{figure}

Note that the two expressions in Eq.~(\ref{Seq:phic}) and Eq.~(\ref{Seq:phi0}) show similar forms except that their degree dependency is inversed.  
The direct comparison between $\phi_c$ and $\phi_0$ is difficult due to degree heterogeneity in $P(k)$, yet the survival probability of species having degree $K$ shows that $\phi_c$ always decreases faster than $\phi_0$ as $\sigma_\xi$ is always smaller than 1.
At $\sigma = 0$, $\phi_c$ and $\phi_0$ for all degree distributions have the same values, $\phi_c = \phi_0 = 1$, and as $\sigma$ increases, they cross over at the triple point $\sigma_t = \sqrt{2K^2/\langle k^2\rangle}$ with $\phi_c = \phi_0 = 1/2$. 
To sum up, the difference between two survival probabilities $\phi_c - \phi_0 = 0$ at $\sigma = 0$ and $\sigma = \sigma_t$, the endpoints of the UFP-UG phase boundary.

If we set the value of $\sigma$ to be sufficiently small ($0<\sigma \ll \sigma_t$), the slope of diversity drop in $\mu$ from $0$ to $\mu_c$ can be classified according to $\gamma$ exponent (see Fig.~\ref{fig:S2}):
\begin{align}
    \frac{\phi_0 - \phi_c}{\mu_c} \approx 
    \begin{cases}
        \displaystyle +\infty \qquad\quad &\mathrm{for}~P(k)\sim k^{-\gamma}~\mathrm{with}~2<\gamma<3~\mathrm{due~to~diverging~}\langle k^2\rangle\\
        \\
        \displaystyle \frac{\langle k^2\rangle}{2K^2}>\frac{2}{3} &\mathrm{for}~P(k)\sim k^{-\gamma}~\mathrm{with}~3<\gamma<4~\mathrm{due~to~diverging~}\langle k^3\rangle\\
        \\
        0 &\mathrm{for}~P(k)\sim k^{-\gamma}~\mathrm{with}~\gamma>4~\mathrm{for}~\sigma\ll\sigma_\xi\leq 1 
    \end{cases}.
\end{align}

%%%%%%%%%%%%%%%%%%%%%%%%%%%%%%%%%55
\section{Numerical Simulation}
\subsection{Simulation details}
The random networks composed of $S$ nodes with a power-law degree distribution for our simulations are generated by assigning the prescribed value $i^{-\alpha}$ to the $i$-th node (the integer node index $i \in [1,S]$). For connections, we select a pair of nodes with probabilities proportional to the product of their assigned values and connect them if they were not already connected.
We repeat this select-and-connect process until the mean degree becomes $K$.
It gives a power-law degree distribution for $\alpha \in (0,1)$, $P(k)\sim k^{-\gamma}$ with $\gamma = 1+1/\alpha$.
This method is called the static model~\cite{goh2001universal}, and the exact form of $P(k)$ is also well-established~\cite{DSLee2004,catanzaro2005analytic}. 
Note that for $\alpha=0$, the model is reduced to the celebrated Erd\H{o}s-R\'enyi random network~\cite{Erdos1959,Gilbert1959} with the Poisson degree distribution.
To investigate the role of the degree heterogeneity, we select three values of $\alpha = 0,\:2/5,\:2/3$, where the corresponding degree distributions are $\mathrm{Pois}(k;K)$, $\sim k^{-3.5}$, and $\sim k^{-2.5}$, respectively. 

Starting from the sampled matrices $(\mathbf{J},\mathbf{A})$ with an initial condition $\mathbf{x}(0)$ uniformly drawn from the interval $[0,1]$ for each element, we numerically integrate Eq.~\eqref{eq:LV} using the Runge-Kutta-Fehlberg method. 
During numerical integration, the species whose abundance goes below the threshold value $10^{-10}$ are considered extinct, so we set their abundance as zero.
Each simulation is terminated when the magnitude of the next updating value $|\Delta x_i| < 5\times 10^{-13}$ for every $i$ or after the time $t=10^6$.
We additionally terminate the simulations when $m>10^5$ due to diverging dynamics.
If a simulation runs over two days, we kill the process and restart with a new initial state.
This case often occurs in the systems belonging to the MA phase, such as quasiperiodic orbits that never converge in a finite time.
The SC equations in MF solutions are numerically solved using the least-squares method.

\subsection{Choice of the mean degree $K$}
\label{sec:ANA}
Selecting the appropriate mean degree $K$ is essential for validating the annealed approximation.
Let us consider a general power-law distribution, $P(k) \sim k^{-\gamma}$.
The expected maximal degree of generated random networks follows $k_\mathrm{max}\sim KS^{1/\omega}$~\cite{catanzaro2005analytic,catanzaro2005generation}, where it takes the natural cutoff $\omega = \gamma-1$, and the annealed network approximation assumes the connection probability satisfies $p_{ij} \approx k_i k_j/(SK) \ll 1$.
In order not to violate this criterion, a strict condition for validity is given by
\begin{align}
    \max_{i,j} p_{ij} \approx \frac{k_\mathrm{max}^2}{SK} \sim KS^{-1+2/\omega} \ll 1.
    \label{Seq:ANA_condi}
\end{align}
This condition cannot be attained for the case where the mean degree is scaled as $K \sim \mathcal{O}(S)$ because it results in $S^{2/(\gamma-1)}\ll 1$, implying the mean degree should be $K\ll S$.
In addition, for $2<\gamma<3$, $K\ll S$ is not enough as $K\ll S^{(\gamma-3)/(\gamma-1)}$.
Thus the annealed network approximation no longer impeccably holds for both cases.

Nevertheless, the condition in Eq.~(\ref{Seq:ANA_condi}) is too tight to select a mean degree in practice.
Considering links between hubs are more sporadic than links between hub-periphery in network realizations, we establish a generous but plausible condition for annealed network approximation as
\begin{align}
    \max_i \bar{p}_i = \max_i \sum_{j\in \mathrm{nn}(i)}  \frac{1}{|\mathrm{nn}(i)|} \frac{k_i k_j}{SK}\ll 1,
    \label{Seq:ANA_naive_condi}
\end{align}
where $\mathrm{nn}(i)$ is a set of the nearest neighbors of $i$-th node.
Instead of considering a connection probability $p_{ij}$ of every possible pair, $\bar{p}_i$ considers representative pairs to be connected on random networks.
We compute the average nearest neighbor degree via the result from Ref.~\cite{catanzaro2005analytic} and choose $K=30$ for our simulations with $S=4000$, which satisfies a criterion for every degree distributions discussed in the main text: $\max_i\,\bar{p}_i \approx 0.59 $, $0.15$, and $0.01$ for degree distributions $\mathrm{Pois}(k;K)$, $\sim k^{-3.5}$, and $\sim k^{-2.5}$, respectively.

\begin{figure}[!h]
    \centering
    \includegraphics[width=0.5\textwidth]{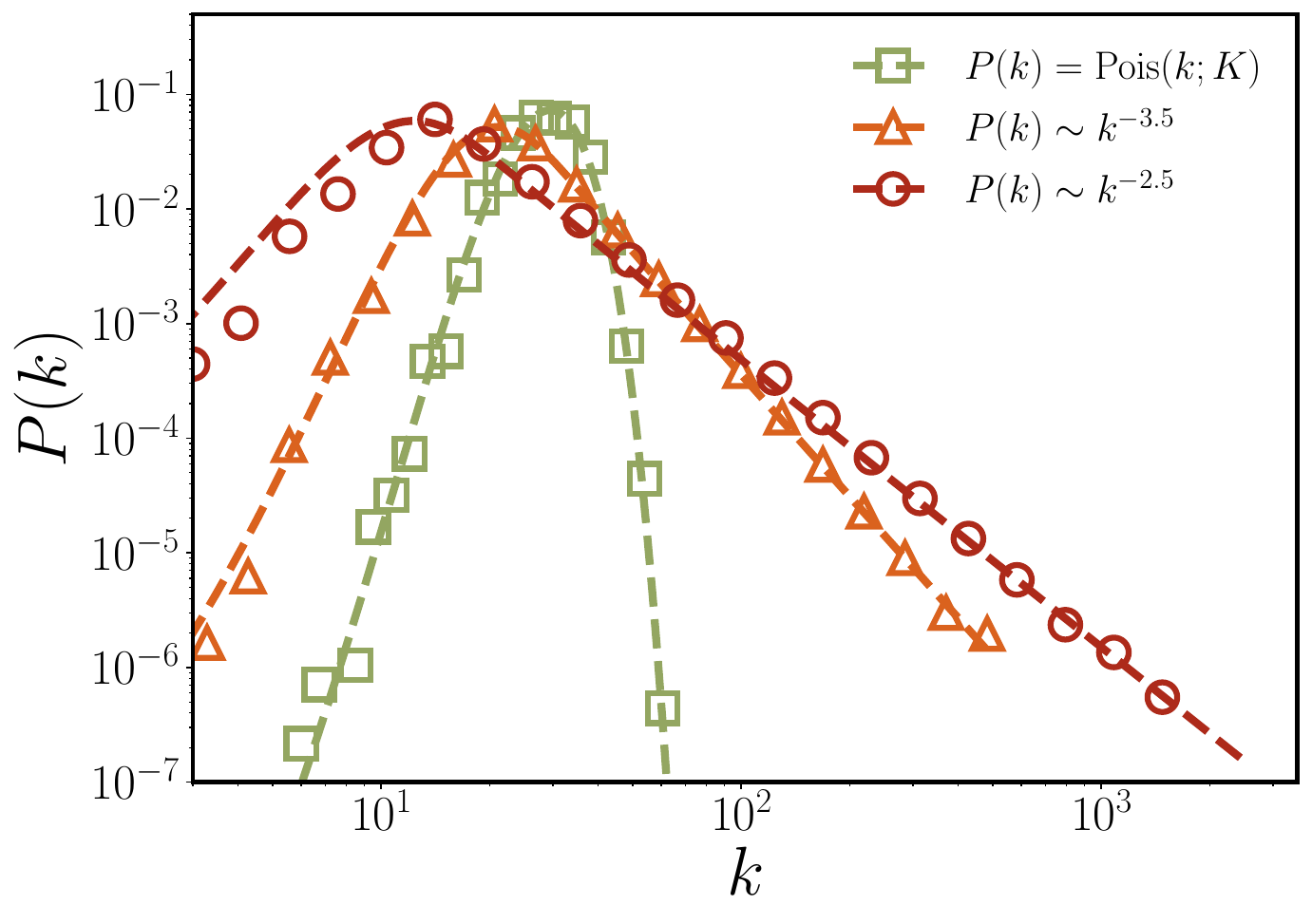}
    \caption{Degree distributions of random networks generated by the static model including the random network with $\alpha = 0$. The curves represent the analytical solution of $P(k)$ in a thermodynamic limit $S\rightarrow \infty$, and $S=4000$ is used for simulations (symbols).}
    \label{fig:S7}
\end{figure}

\subsection{Additional simulation results}
\begin{figure}[!h]
    \centering
    \includegraphics[width=\textwidth]{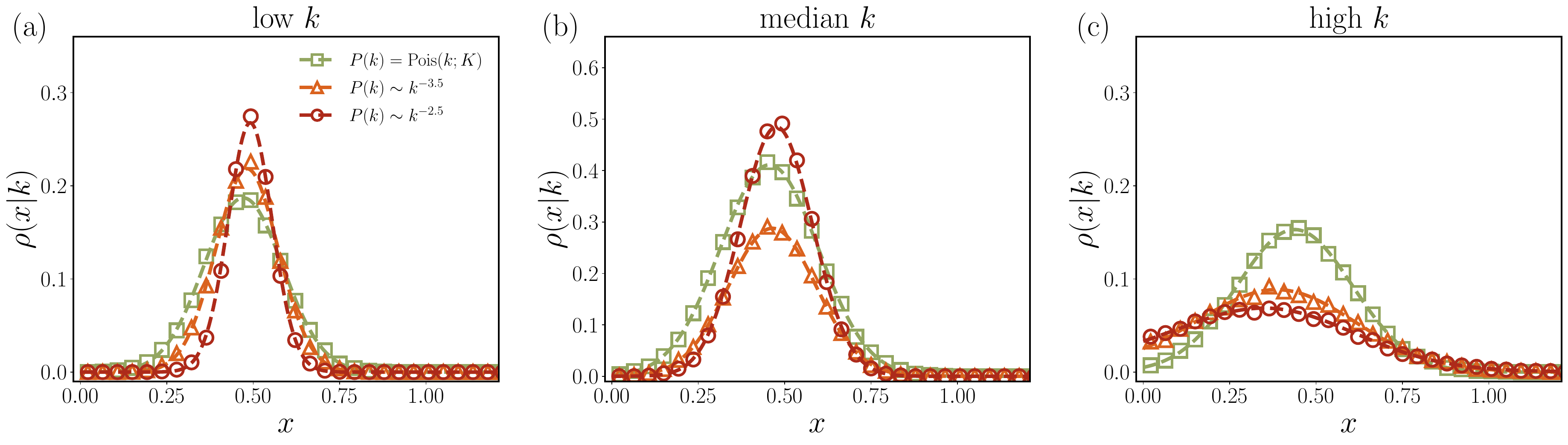}
    \caption{The conditional abundance distribution $\rho(x|k)$ at $(\mu,\sigma) = (0.1,0.3)$ with different degrees. We coarse-grain the simulation data with cumulative degree distribution $C(k) = \sum_{k'=1}^k P(k')$, and the degrees satisfying (a) $C(k)\leq 0.05$ for the low degree, (b) $0.475\leq C(k)\leq0.525$ for the median degree, and (c) $C(k) \geq 0.95$ for the high degree are used.}
    \label{fig:S3}
\end{figure}

To support the validity of our MF method, we compare the MF prediction and simulation results of the conditional abundance distribution $\rho(x|k)$ for various $k$ in Fig.~\ref{fig:S3}.

\begin{figure}[!h]
    \centering
    \includegraphics[width=\textwidth]{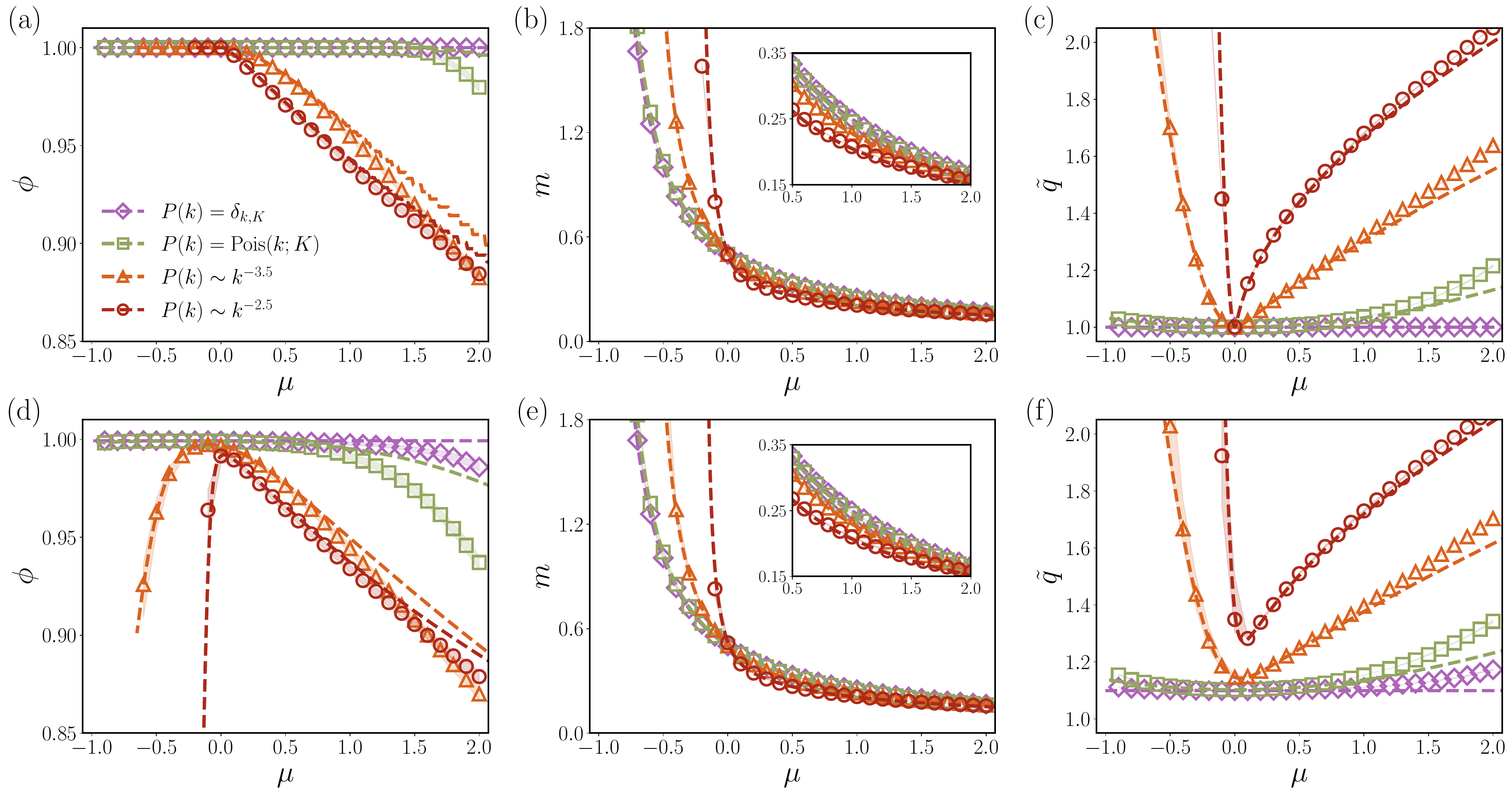}
    \caption{Three order parameters $\phi$, $m$, and $\tilde{q}$. The case for homogeneous interaction strength $(\sigma = 0)$ is drawn in the panels (a), (b), and (c). As the MF solution (dashed line) predicted, a stable fixed point is feasible for $\mu<0$, i.e. $\phi = 1$. The panels (d), (e), and (f) show the results with heterogeneous interaction strength $\sigma = 0.3$. In cooperative communities, the survival probability $\phi$ drops steeper on networks with stronger heterogeneity. Not having enough link numbers (small $K$) causes a discrepancy between the MF solution and simulation results.}
    \label{fig:S4}
\end{figure}

Figure~\ref{fig:S4} shows $\phi$, $m$, and $\tilde{q}$ on networks with different degree distributions.
With identical interactions, $\sigma = 0$, all species survive in a cooperative community.
At the UG critical point $(\mu_c, \sigma_c) = (-K^2/\langle k^2\rangle, 0)$, $m\sim{(\mu-\mu_c)}^{-1}$ always diverges by definition, whereas $\tilde{q} = K\langle k^3\rangle/\langle k^2\rangle^2$ diverges depending on the exponent $\gamma$ of the power-law degree distributions $P(k)\sim k^{-\gamma}$. 
When $2<\gamma<4$, $\tilde{q}$ diverges. 
Since the width of reduced abundance distribution is proportional to $\sigma\sqrt{\tilde{q}}$ as shown in Eq.~(\ref{Seq:xtilde}), the structural heterogeneity drags the survival probability $\phi=\langle \Theta(x)\rangle = \langle \Theta(\tilde{x})\rangle$ down when it combined with heterogeneous interaction strength.

\begin{figure}
    \centering
    \includegraphics[width=0.5\textwidth]{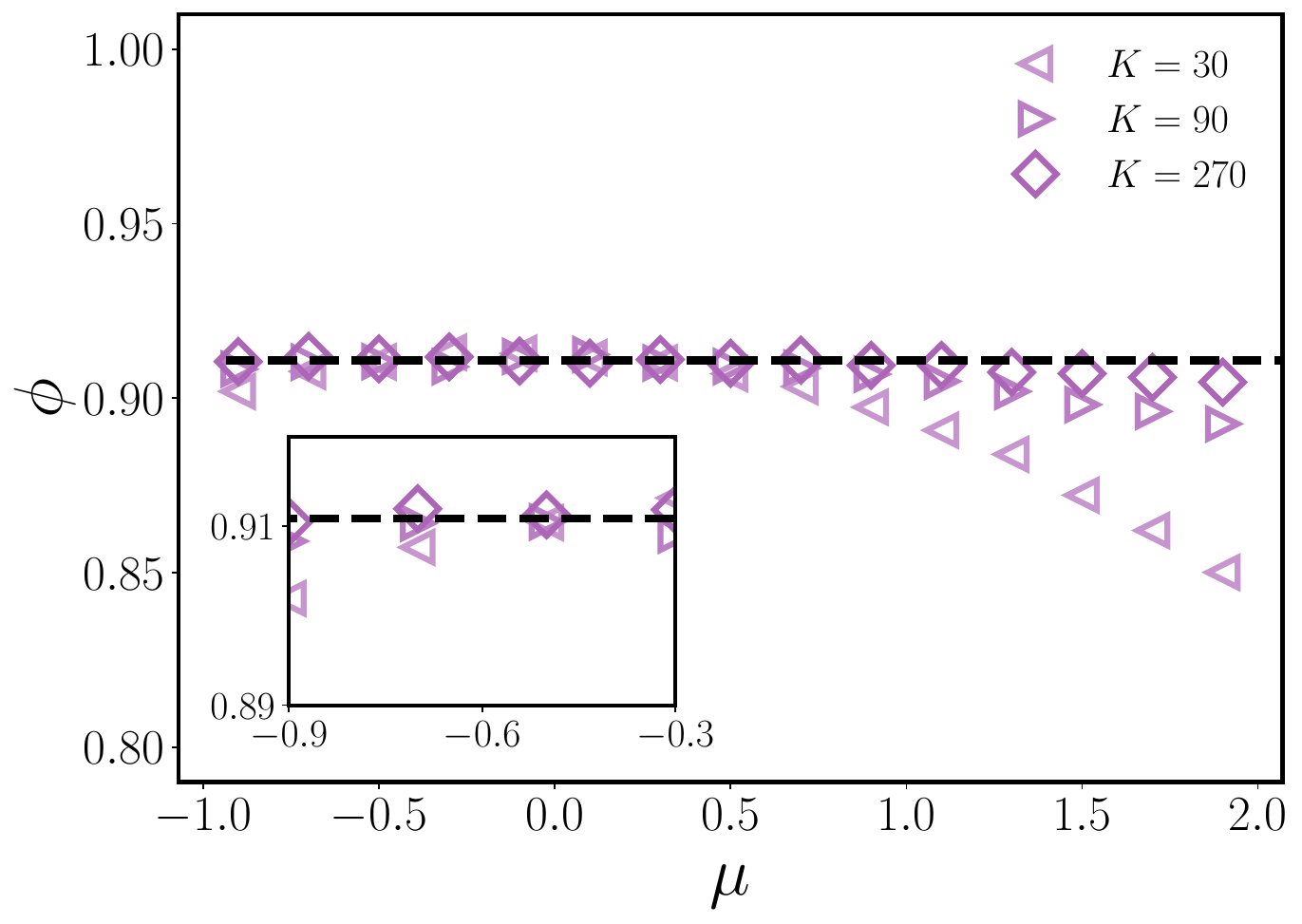}
    \caption{The survival probability $\phi$ on regular random networks with $\sigma = 0.6$. A large gap exists between MF prediction and simulation results for $K=30$ and shows a drop in $\phi$ as $|\mu|$ increases, which is not expected in the MF solution. The simulation results approach the MF solution as $K$ increases, guaranteeing the MF argument is valid for the large $K$ limit.}
    \label{fig:S5}
\end{figure}

Another noteworthy feature in Fig.~\ref{fig:S4} is that the gap between the theoretical prediction and numerical simulations widens as $|\mu|$ increases.
The mismatch between them is also detected under higher heterogeneity in interaction strength.
For instance, for the regular random networks with $\sigma=0.6$ in Fig.~\ref{fig:S5}, the survival probability $\phi$ should be constant in $\mu$ with our analysis but lesser species survived in our simulation results for $K=30$.
This violates our hypothesis: the complex of structural and strength heterogeneities triggers the decreasing diversity under stronger cooperation.
It stems from insufficiently large values of $K$.
Back to the derivation in Eq.~(\ref{Seq:Korder}), the MF solution is exact in the ``dense'' limit $K\rightarrow \infty$ (more precisely, $1\ll K \ll S$, or $K\sim S^\theta$ with $0<\theta<1$, and this link-species power-law relationship is previously reported in ecology~\cite{brose2004unified,garlaschelli2003universal}). 
When $K$ is not large enough, it is not allowed to dismiss the higher-order terms such follow $\mathcal{O}(K^{-2})$ in Eq.~(\ref{Seq:Korder}), making the significant deviation.
We validated this finite $K$ effect by investigating the survival probability in denser networks, specifically with $K=90$ and $K=270$, demonstrating convergence to our MF solution.

\begin{figure}%[!h]
    \centering
    \includegraphics[width=0.5\textwidth]{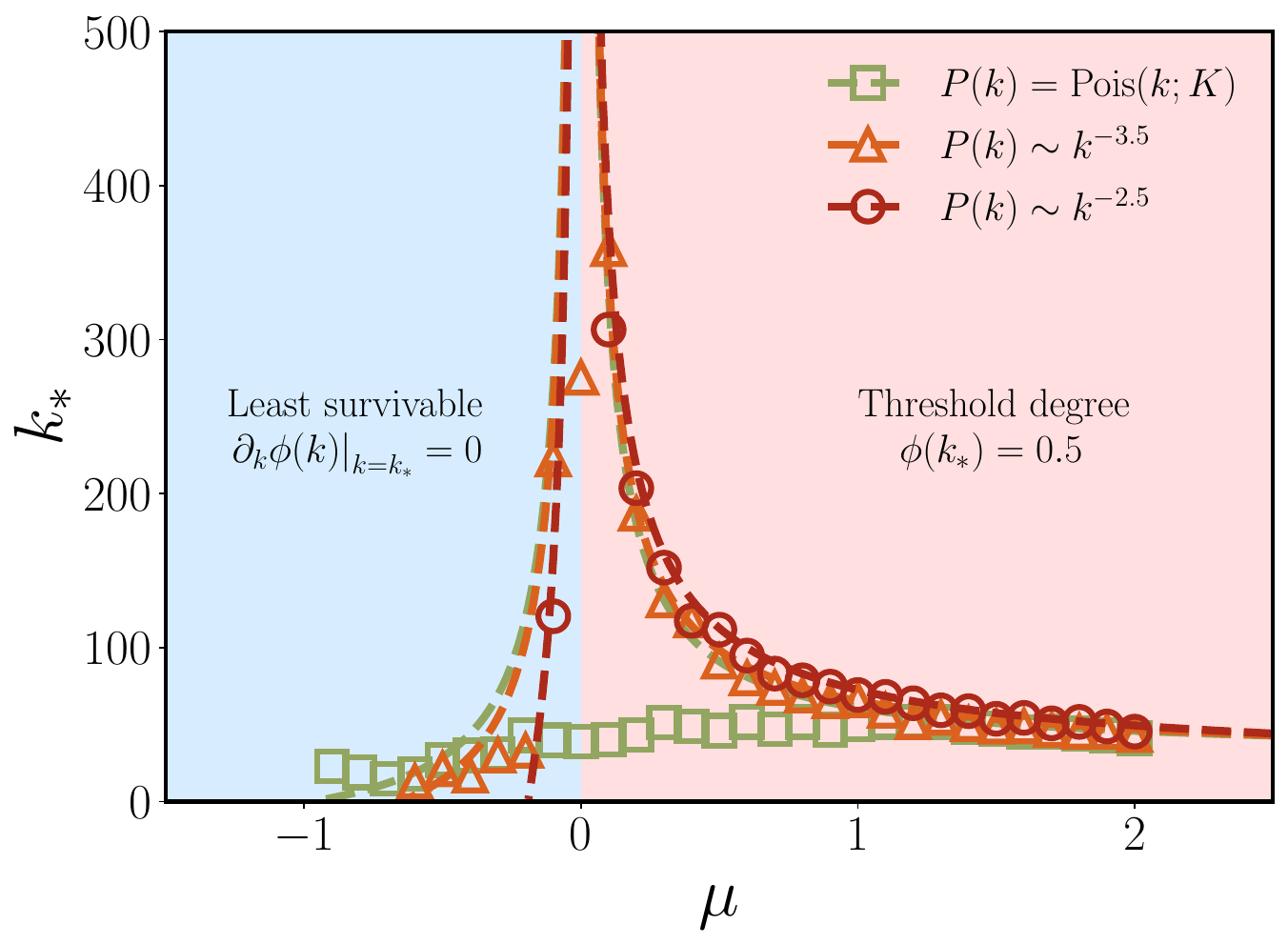}
    \caption{The characteristic degree $k_* = \lambda K /(|\mu|m)$ at $\sigma = 0.3$ as a function of $\mu$. The $k_*$ acts as a threshold degree that the species with a degree larger than $k_*$ almost vanished in competitive communities, while the species with degree $k_*$ is the least survivable in cooperative communities. The mismatch between MF prediction and simulations for the Poisson distribution is caused by the finite size of $S$ (or the finite number of samples, equivalently), where its expected maximal degree is $k_{\mathrm{max}} \approx 58.12$ for $S = 4000$ and $K=30$.}
    \label{fig:S6}
\end{figure}

The characteristic degree $k_* = \lambda K/(|\mu|m)$ is plotted as a function of $\mu$ in Fig.~\ref{fig:S6}, showing a monotonic decrease with respect to $|\mu|$. 
This deduction may not be immediately apparent due to the trend of $m$.
The species with degree $k_*$ have minimal survival probability in cooperative communities, according to $\left.\partial_k\Delta(k)\right|_{k=k_*} = 0$ for $\mu<0$. 
Due to diverging $m$ at $\mu = \mu_c$, $k_*$ in cooperative communities sharply decreases to zero as $\mu \rightarrow \mu_c$, indicating that peripheral species are more likely to extinct under strong cooperation.
On the other hand, $\Delta(k_*) = 0$ in competitive communities, informing that at best half of species survive if their degrees exceed $k_*$. 
This informs that the peripheral species are more likely to survive under strong competition.
In Fig.~\ref{fig:S6}, the characteristic degrees extracted from the simulation results via $k_* = {\arg\!\min}_k\:\phi(k)$ for $\mu<0$ and $k_* = {\arg\!\min}_k \:|\phi(k)-0.5|$ for $\mu>0$ show good agreements with MF solutions $k_* = \lambda K /(|\mu|m)$ for two power-law degree distributions.
The reason why the case for the Poisson distribution does not follow the MF prediction is a finite-size effect, where a degree larger than $k_{\max} \approx 58$ hardly formed in a single realization of $\mathbf{A}$ with $S=4000$ and $K=30$~\cite{briggs2009note}. 

\subsection{The measurements for phase discrimination}
We introduce the indices $\mathcal{I}_1$ and $\mathcal{I}_2$ to distinguish which phase the system belongs to from the numerical simulations.
Presuming the system has diverging abundance when $m$ exceeds $10^5$, the index $\mathcal{I}_1$ measures how many realizations of $(\mathbf{J}_\alpha,\mathbf{A}_\alpha)$ are at the UG phase:
\begin{align}
    \mathcal{I}_1 \equiv \frac{1}{N_\mathrm{sample}}\sum_{\alpha=1}^{N_\mathrm{sample}} \Theta\left(m_\alpha - 10^5\right),
\end{align}
where $N_\mathrm{sample}$ is the number of sampled random matrices $(\mathbf{J}_\alpha,\mathbf{A}_\alpha$) and $m_\alpha = m(\mathbf{J}_\alpha,\mathbf{A}_\alpha)$ is $m$ from $\alpha$-th sample.

For the MA phase, we consider the correlation function between two systems $\mathbf{x}_1(t)$ and $\mathbf{x}_2(t)$ with identical sampled matrices $(\mathbf{J},\mathbf{A})$ but different initial conditions,
\begin{equation}
\begin{aligned}
d^2(t,t') &\equiv S^{-1}\sum_i \frac{k_i}{K}\left\langle [x_{1,i}(t) - x_{2,i}(t)][x_{1,i}(t') - x_{2,i}(t')]\right\rangle_0\\
&= q_{11}(t,t') + q_{22}(t,t') - q_{12}(t,t') - q_{21}(t,t'),
\label{Seq:d2}
\end{aligned}
\end{equation}
where $q_{\beta\gamma}(t,t') = \langle k x_\beta(t) x_\gamma(t') \rangle /K$ with $\beta,\gamma \in \{1,2\}$.
The equitemporal correlation in the long time limit $d^2 \equiv \lim_{t\rightarrow \infty} d^2(t,t)$ gives a mean-squared weighted Euclidean distance between two stationary states with symmetric order parameter $q_{\beta\gamma} = \lim_{t\rightarrow\infty} q_{\beta\gamma}(t,t)$.
Starting from different initial configurations $\mathbf{x}_1(0)$ and $\mathbf{x}_2(0)$, both will converge to an identical stationary state $\mathbf{x}_1 = \mathbf{x}_2 = \mathbf{x}^*$ at the UFP phase, resulting $d^2 = 0$.
Once there are multiple equilibria, their final configurations would be different even if their macroscopic quantities are equal $q_{11} = q_{22}> q_{12}$, leading to $d^2>0$.
When their distance $d^2$ is larger than $5\times 10^{-3}$, we regard that two states are distinct. 
The index $\mathcal{I}_2$ measures how many realizations of $(\mathbf{J}_\alpha, \mathbf{A}_\alpha)$ are at the MA phase:
\begin{align}
    \mathcal{I}_2 \equiv \frac{1}{N_\mathrm{sample}} \sum_{\alpha=1}^{N_\mathrm{sample}} \Theta \left(\max_{\beta,\gamma}  d^2_{\alpha}(\mathbf{x}_\beta,\mathbf{x}_\gamma) - 5\times10^{-3} \right),
\end{align}
where $d^2_\alpha(\mathbf{x}_\beta,\mathbf{x}_\gamma)$ is $d^2$ between two copies $\mathbf{x}_\beta$ and $\mathbf{x}_\gamma$ from $\alpha$-th sample.

In addition, we can relate the correlation function in Eq.~(\ref{Seq:d2}) to chaotic behavior.
Let us choose $\mathbf{x}_1(t)=\mathbf{x}^*$ as the MF stationary solution in Eq.~(\ref{Seq:fp}) and consider a small perturbation as $\mathbf{x}_2(t) = \mathbf{x}^* + \epsilon\,\delta\mathbf{x}(t)$ with ${\langle \delta x_i(t) \rangle}_0 = 0$.
In this case, we obtain the correlation function as $d^2(t) = \delta q(t) = \epsilon^2\,\langle k\,{\delta x(t)}^2\rangle/K$, where its integrated version dealt with the transition to the MA phase in Sec.~\ref{sec:MA_phase}.
Therefore, one could use the maximal Lyapunov exponent with weighted Euclidean distance as another candidate for $\mathcal{I}_2$,
\begin{align}
    \ell_{\mathrm{max}} = \lim_{\substack{t \rightarrow \infty \\ \epsilon \rightarrow 0}} \frac{1}{2t} \log \frac{d^2(t)}{d^2(0)} = \lim_{\substack{t \rightarrow \infty \\ \epsilon \rightarrow 0}} \frac{1}{2t} \log \left[\frac{\langle k\,\delta x(t)^2\rangle/K}{\langle k\,\delta x(0)^2\rangle/K}\right].
\end{align}
The maximal Lyapunov exponent makes the phase discrimination more systematic, requiring only the assessment of its sign, given that the transition occurs precisely at $\ell_\mathrm{max} = 0$ (see Ref.~\cite{schuecker2018optimal} for a decent solving technique for the Lyapunov exponent).

%%%%%%%%%%%%%%%%%%%%%%%%%%%%%%%%%%%%%%%%%%%
%\bibliographystyle{apsrev4-1}
%\bibliography{reference.bib}
%merlin.mbs apsrev4-1.bst 2010-07-25 4.21a (PWD, AO, DPC) hacked
%Control: key (0)
%Control: author (72) initials jnrlst
%Control: editor formatted (1) identically to author
%Control: production of article title (-1) disabled
%Control: page (0) single
%Control: year (1) truncated
%Control: production of eprint (0) enabled
%
%%%%%%%%%%%%%%%%%%%%%%%%%%%%%%%%%%%%%%%%%%%